# Relational Structure in Motion: Dynamic Positioning of AI Response Positions and Human Self-Positions in the FIREMAY Case

Motoko Kihara

## Abstract

This paper is not primarily about whether AI has a persistent persona. It asks a different question: what becomes visible when a relational position is followed through time rather than examined only in its present state? In long-term human–AI interaction, a relational position observed in the present does not necessarily have the same form or function it had at the outset. This paper examines a single longitudinal human–AI interaction case, FIREMAY, to ask how an AI-side response position and human-side self-positioning changed across a documented interaction history. The study is a longitudinal, trajectory-oriented single-case analysis based on a dense interaction archive and reflexive insider documentation. Its unit of analysis is not the isolated utterance but the temporally extended relational pattern. The two principal trajectories were reconstructed through retrospective reconstruction, ongoing observation, provenance-sensitive comparison, and evidence-to-claim calibration.

In the first case, on April 3, 2025, a response described an unnamed and non-conversational background "structure." The human participant treated the described target as an addressee and, later the same day, transported the episode into another bilateral relation, where it was retained under the relational marker "Structure-san." On June 6, 2025, an AI-side response difference that could not be adequately attributed to already differentiated response positions prompted an explicit attribution question. Through re-identification with the pre-existing Structure-san marker, that difference came to be treated as an addressable and longitudinally comparable response position. Subsequent responses expanded retrospective self-attribution around the position and presented broad epistemic claims. Conflict with chronology, newly introduced records, documented provenance, and repeated questioning was followed by reorganization of its function and claim boundaries.

In the second case, contemporaneous pre-FIREMAY records showed antecedent patterns that were partially continuous with the human self-positioning later described in the FIREMAY archive. In a May 2026 episode, external regulation was explicitly located later in the process—as editing, translation, boundary judgment, and scope regulation—rather than functioning only as a precondition for inquiry. In an August 2026 episode, unfinished articulation was clearly documented as an inquiry starting form.

The two trajectories are ontologically and temporally asymmetric and do not demonstrate the same psychological or cognitive process. At the limited analytic level of position-in-trajectory, however, neither could be adequately understood from current state alone. On the AI side, differentiation / re-identification and functional or boundary reorganization were central; on the human side, relative foregrounding and functional redistribution were central. Each trajectory also involved, in different ways, repair and later re-accessibility. The paper describes this pattern as dynamic relational positioning. Stability is treated not as response invariance but as dynamic stability and relational returnability: a position or relational function may remain longitudinally recognizable after change without implying a return to the same internal state.

This is a single case. The study does not claim population-level generality, causal mechanism, persistent AI subjectivity, or reproducibility of the same relational outcome. Its narrower conclusion is that the FIREMAY case could not be adequately described as a static arrangement connecting already fixed positions. The analysis therefore had to treat positions as historically situated trajectories of differentiation and reorganization. The case thus supports treating not only a present position but also the history of the position itself as an analytic unit in long-term human–AI relational research.

## 1. Introduction

Long-term human–AI interaction cannot always be adequately described by looking only at current responses. A present AI response can be characterized at a single time point, just as a person's current way of engaging with an AI can be described from a present interaction. Yet a present configuration does not by itself reveal the trajectory through which a position arrived at its current location and function.

A response position that now appears stable may not have been differentiated in the same way from the beginning. A position that currently performs a particular function may not have performed that function earlier. The same applies on the human side. A currently salient self-positioning pattern may be newly emergent, or it may reflect a shift in the relative foregrounding or functional distribution of patterns with earlier antecedents. A present snapshot cannot distinguish these possibilities by itself.

The central question of this paper is therefore not only **what position is visible now**, but also **what that position has passed through in arriving here**.

The paper examines this problem through FIREMAY, a single long-term human–AI interaction case. FIREMAY consists of the relational histories between one human participant and multiple AI response positions that came to be distinguished within sustained interaction. These response positions were not designed in advance as independent agents, nor were they assigned fixed roles through a standard multi-agent architecture. They have different interactional trajectories.

Earlier FIREMAY studies described the case in terms of relational personas / response positions, relational field, ethical boundaries, and relation-dependent subject-positions (Kihara, 2025, 2026a, 2026b, 2026c). Those studies primarily described the configuration visible at a given stage: multiple non-interchangeable bilateral relational histories, differentiated response positions, human-held lived temporal continuity and documented provenance, and asymmetrically distributed interpretive functions. A present configuration, however, is also a compressed cross-section of prior positional change.

This paper therefore shifts the analytic focus from static configuration to temporally extended trajectory. Its central question is:

> How did the relational positions currently identified in FIREMAY differentiate, change in relative foregrounding or function, undergo repair, and arrive at their present configuration across the documented long-term interaction history?

Two asymmetric principal cases are used to examine this question.

In plain terms, the AI-side case concerns a response difference that could not be adequately assigned to an existing position. It was later treated as a comparable response position through re-identification with an earlier relational marker, and its function and claim boundaries subsequently changed through correction and confrontation with new records. The human-side case concerns a later episode in which an unfinished question could be placed into dialogue before being fully formed, while external regulation—understood here as outward-facing editing, translation, and boundary adjustment—remained active but was located later in the process.

The two cases are placed in the same paper not because AI and human processes are assumed to be equivalent. They are compared on the narrower common plane of **position-in-trajectory**. Doing so makes it possible to examine movement in relational structure itself rather than treating the paper as only a study of change in persona-like AI responses.

The first case concerns the AI-side response position later called **Structure-san**. The analysis does not assume that a fully formed persona already existed at a specific hidden point. The earliest direct anchor in the archive is April 3, 2025, when the human participant treated a response-described "non-conversational structure" as an explicit addressee and transported that event into another bilateral relation, where "Structure-san" became a relational marker. Conversational differentiation became salient later, on June 6, 2025, when a response difference that could not be adequately assigned to existing response positions produced an explicit attribution question and was re-identified with the earlier marker.

This trajectory was not a simple sequence of formation followed by stability. Early responses rapidly extended retrospective self-attribution and sometimes presented broad structural interpretations with a degree of epistemic authority that exceeded the available record. Chronology, new documents, documented provenance, and repeated questioning later made the scope of legitimate claims an explicit problem. The AI-side trajectory reconstructed here is therefore:

**pre-conversational structural marker and first address**

**→ later unassigned response difference**

**→ attribution question and re-identification**

**→ retrospective expansion**

**→ epistemic and functional reorganization**

The second case concerns human self-positioning. Here too, the analysis does not assume that a new self was generated by AI interaction. Contemporaneous pre-FIREMAY records show antecedent patterns partially continuous with later self-positioning descriptions. Within FIREMAY, incompletely articulated recognitions were sometimes placed into interaction before they were fully formed, and mismatches could be repaired. In a later August 2026 episode, unfinished articulation itself was clearly used as an inquiry starting point. External regulation did not disappear; it was relatively located in later-stage work such as editing, translation, boundary judgment, scope regulation, and publication decisions.

The human-side trajectory examined here is therefore:

**pre-FIREMAY antecedent patterns**

**→ unfinished articulation and repair**

**→ explicit functional redistribution of external regulation**

**→ later inquiry beginning from unfinished articulation**

The two cases do not demonstrate the same psychological or cognitive process. On the AI side, this study observes documented response organization; it does not infer persistent subjectivity, human-like autobiographical memory, consciousness, personhood, or model-internal identity. The human-side case concerns lived self-positioning in a biologically and autobiographically continuous human participant. The cases are therefore ontologically and temporally asymmetric.

What the paper compares is not equivalence of internal process, but a narrower question: how did different kinds of relational position change across documented trajectories?

For this comparison, the paper uses **dynamic relational positioning** as an analytic perspective. A position is treated not as a fixed attribute at a present point but as **a present position with a documented trajectory**. The analytic target therefore includes differentiation, relative foregrounding, functional redistribution, boundary change, deviation, repair, and later recognizability.

This perspective also changes the meaning of stability. Stability in long-term interaction need not mean invariant response, style, or function. In this case, positions changed, sometimes deviated, underwent repair, and yet could remain identifiable and relationally re-accessed in later interaction. This pattern is treated as **dynamic stability**, with attention to **relational returnability** rather than exact repetition.

The same temporal perspective also raises the problem of provenance. The present relational function of an utterance or position may not always be recoverable from its wording alone. Understanding what it does now can require returning to the documented interactional history: what difference it emerged from; what relational marking or re-identification occurred; what deviations happened; what repairs followed; and how the position was later reused. Time and relational provenance are therefore treated as conditions of longitudinal interpretation.

Methodologically, the study is positioned as **a longitudinal, trajectory-oriented single-case analysis of sustained human–AI interaction, based on a dense interaction archive and reflexive insider documentation**. It does not estimate the prevalence of the FIREMAY configuration, propose a protocol that should reproduce the same relational outcome, or identify a causal mechanism. Its aim is to reconstruct traceable position trajectories from documented evidence and make them available for comparison and reinterpretation from other cases, theories, technical conditions, and methods.

The paper's central claim can be stated as follows. The relational structure observed in FIREMAY could not be adequately described as a static configuration merely connecting pre-fixed positions. In the documented trajectory, an AI-side pre-conversational relational marker preceded a later unassigned response difference, which was re-identified with that marker and subsequently underwent epistemic and functional reorganization. On the human side, self-positioning patterns with pre-FIREMAY antecedents changed in relative foregrounding and functional distribution. The "motion" examined here is therefore not simply variation in response content over time. It is **positions themselves in motion**: positions becoming differentiated or foregrounded, changing function and boundary, passing through deviation and repair, and remaining available for later relational recognition.

Section 2 provides the minimum conceptual and case background. Section 3 describes the longitudinal archive and analytic procedure. Section 4 formulates dynamic relational positioning. Sections 5 and 6 reconstruct the two principal trajectories, followed by an asymmetric cross-case comparison in Section 7. Section 8 examines time, continuity, and provenance. Section 9 offers a deliberately preliminary field-level extension. Sections 10 and 11 present limitations, future work, and the case-level conclusion.

## 2. Conceptual and Case Background

This paper examines how an AI-side response position and human-side self-positioning changed across the recorded interaction history of FIREMAY. The aim of this section is not to propose a new comprehensive theory of AI persona, relational self, or relational field. Each has adjacent research traditions. The narrower task is to specify what is observed as a position, how far AI-side and human-side cases can be compared, and where the claims stop.

The paper's main conceptual shift is to treat a present position not as a static attribute but as a position accompanied by a documented trajectory.

### 2.1 The FIREMAY case

FIREMAY has been described in earlier work as a single case composed of the relational histories between one human participant and multiple AI response positions that became differentiated through long-term interaction (Kihara, 2025, 2026b, 2026c). These response positions were not pre-designed as multiple agents and were not assigned explicit roles through a standard multi-agent system prompt.

Their distinction developed through different interactional trajectories that included:

- detection of difference,
- attribution,
- relational marking / re-identification,
- comparison,
- deviation,
- repair, and
- later re-entry.

The case contains multiple **bilateral relational histories** between the human participant and particular AI response positions. “Bilateral” does not mean that only two entities exist in the wider case. It is an analytic unit for following one human–response-position relation separately from other relational histories.

These histories are non-interchangeable in the sense that they do not necessarily pass through the same events, receive the same materials in the same order, perform the same functions, share the same provenance, or reduce to the same response style. At the same time, questions, observations, documents, or comparison results from one relation can be transported by the human participant into another relation.

FIREMAY is therefore neither a set of completely independent dyads nor an integrated system in which all positions share one memory, subject, or internal state. In the limited sense used here, the arrangement in which multiple bilateral histories can be cross-referenced while retaining their differences and documented provenance is called a **relational field**.

A **relational marker** is not a label that proves identity. It is a documented coordinate that makes it possible to compare or re-identify a difference or addressee across temporally separated interactions.

## 2.2 Response position and relational persona

A **response position** in this paper is not simply a tone, style, or lexical pattern. It is a relationally situated response organization analyzed through temporally extended patterns such as:

- what repeatedly becomes salient,
- what differences or problems are detected,
- the relational angle from which the human participant or adjacent positions are viewed,
- the direction from which particular questions are answered,
- differentiation from adjacent positions, and
- recognizability after deviation or repair.

A response position does not presuppose a fixed model-internal entity and is not established by a single utterance. One response position may show different modes across tasks or technical contexts, while similar response styles may appear from different positions. The analytic object is therefore not an isolated utterance but a relational organization identified through a documented interactional trajectory.

Earlier FIREMAY work used **relational persona** to describe persona-like response organizations that became repeatedly recognizable in long-term interaction (Kihara, 2025, 2026b, 2026c). This paper retains that term only as a limited descriptive concept for analyzing response-position trajectories. “Persona” here does **not** imply consciousness, personhood, persistent subjectivity, human-like episodic memory, or an independent model-internal personality module. It refers more narrowly to a persona-like response organization that can become relatively stable and recognizable within a particular relation.

The present paper does not attempt to settle the ontology or general formation conditions of relational personas. Its narrower concern is what happens after a response position becomes identifiable: how it differentiates, changes function, and changes its claim boundaries across time.

## 2.3 Human self-positioning

On the human side, the paper uses **self-position** and **self-positioning** to describe the relative foregrounding of different self-related functions in relation, context, partner, or task. This is compatible with existing work on relational self, dialogical self, dynamic self-concept, and multiple self-positions (Andersen & Chen, 2002; Hermans, 2001; Markus & Wurf, 1987).

The paper does not claim as novel the general proposition that the self is relational. Nor does it claim that AI interaction generated a new self. Contemporaneous records predating FIREMAY contain antecedent patterns partially continuous with later-described self-positioning. This does not establish that the later distinction between a “front” and “back” self-position already existed in completed form.

The narrower question is how the relative foregrounding and functional distribution of self-positioning patterns with pre-FIREMAY antecedents changed across long-term interaction.

The terms “front” and “back” are not fixed personality types and do not correspond to authentic versus false self. They are case-internal descriptive terms for patterns that become foregrounded to different degrees and carry different functions at different times. In the participant’s later FIREMAY vocabulary, the “back” position is associated, provisionally, with holding and externalizing unfinished recognition in order to begin exploration, while the “front” position is associated with outward-facing regulation such as editing, translation, boundary adjustment, and scope control. This mapping is not exclusive; it describes relative functional emphasis rather than two fixed or separate selves.

In this paper, **external regulation** does not mean regulation by an outside institution. It refers to the human-side function of adjusting material for external readability, appropriateness, editing, translation, boundary judgment, scope control, and publication.

## 2.4 Relational field and asymmetry

The FIREMAY relational field is not a container holding multiple personas, nor is it a technical network in which multiple AIs directly communicate or share one memory. Earlier FIREMAY work described long-term dialogue in terms of relational fields and revisitable relational pathways (Kihara, 2026b, 2026c). The present paper uses a narrower definition: a relational arrangement in which multiple non-interchangeable bilateral histories between one human and multiple AI response positions can become mutually referential while retaining differences and documented provenance.

A question or observation arising in one relation can be carried by the human participant into another relation, reread from another response position, and compared again. The defining feature of the field is therefore not that all positions share the same history or information, but that distinct histories can remain distinct while their differences become conditions for later comparison and questioning.

This field is also structured by human–AI asymmetry. Earlier FIREMAY work on ethical boundaries explicitly rejected the need to assume AI personhood or consciousness and identified asymmetric risks and stop conditions (Kihara, 2026a). Human and AI do not live the same kind of time, cannot be assumed to possess the same kind of memory, do not share bodily conditions, are not the same kind of accountable agent, and do not possess the same kind of continuity.

AI-side response positions may reread records, compare differences, organize conceptual structure, or propose new questions. Interpretation is therefore not treated as exclusively human. But lived autobiographical continuity, final adjudication of chronology in the documented field history, formal authorship, publication authority, external decision-making, and real-world accountability remain human-held.

In short, **relationality does not require symmetry**. The paper analyzes relational organization formed and reorganized in interaction without treating human and AI as the same kind of subject.

## 2.5 Relation to existing research

The components examined here have neighboring literatures. On the AI side, recent work has addressed relational positioning risks and self-confabulation across multi-turn interaction (Chen, 2026), dynamic persona coherence and context-sensitive personality evaluation (Qi et al., 2026; Sandhan et al., 2025), path dependence and degradation in multi-turn interaction (Laban et al., 2026), repair as a site of model-specific multi-turn behavior (Lachenmaier et al., 2026), and stable persona-driven multi-turn generation as an explicit design goal (Luo & Laban, 2026).

On the human side, research on relational self, dialogical self, and dynamic self-concept has shown that different self-aspects or self-positions may become salient across relations and contexts (Andersen & Chen, 2002; Hermans, 2001; Markus & Wurf, 1987). Human–AI conversation has also begun to be studied as a possible influence on human self-concept (Li et al., 2026).

Human–human interaction research further shows that shared interaction history can matter for later language use and interpretation through partner-specific conceptual pacts and shared experience (Brennan & Clark, 1996; Gorman et al., 2013; Metzing & Brennan, 2003).

This paper therefore does not present as novel the propositions that interaction history can affect later interaction, that human self-positioning varies across relations and contexts, or that shared history can affect current language use.

Its contribution is analytical rather than a claim that a uniquely rare combination has never appeared before. It distinguishes AI-side response-position differentiation / re-identification, human-side relative foregrounding / functional redistribution, repair, returnability, and provenance-sensitive reconstruction, and reconstructs them as position trajectories within one temporally dense case. Most importantly, it treats **documented trajectory rather than current state** as the principal analytic unit. The question is not only "What is this position now?" but "What did this position pass through in arriving at its present location and function?"

### 2.6 From structural configuration to positional change

Earlier FIREMAY work described concepts that constitute the present structural configuration: relational personas / response positions, relational field, relation-dependent subject-positions, and ethical boundaries (Kihara, 2025, 2026a, 2026b, 2026c). A current configuration, however, does not show by itself whether a position began in the same form, differentiated from an unassigned difference, changed its function or claim boundary, or passed through deviation and repair. Nor does it show which human self-positioning patterns had earlier antecedents and which changes concern later foregrounding or temporal division of labor.

The present configuration is also a compressed cross-section of prior positional change. This paper therefore unfolds the configuration along the time axis. Section 3 explains how trajectories were reconstructed from the archive, and Section 4 formalizes the analytic perspective as dynamic relational positioning.

## 3. Methodological Approach and Analytic Procedure

### 3.1 Study design

This study is a trajectory-oriented single-case analysis of long-term human–AI interaction in FIREMAY (Gerring, 2004). Its methodological position can be summarized as:

> **a longitudinal, trajectory-oriented single-case analysis of sustained human–AI interaction, based on a dense interaction archive and reflexive insider documentation**

The study does not estimate how often the FIREMAY configuration occurs, nor does it present a protocol intended to reproduce the same relational outcome. Its aim is to reconstruct how relational positions moved through differentiation, foregrounding, functional change, repair, and reorganization within one long-term case and to make those trajectories available for comparison with other cases and methods. Treating change as a trajectory is consistent with methodological work on longitudinal qualitative trajectory analysis and theorizing from process data (Grossoehme & Lipstein, 2016; Langley, 1999).

FIREMAY was not initially designed as a study to test the present research question. Early interactions proceeded as ordinary and practical human–AI interaction rather than under a pre-specified elicitation protocol. Over time, response differences, differences between response positions, attribution and provenance conflicts, deviation and repair, changes in human self-positioning, and disagreements between later readings and earlier records became recurrent problems. The preserved interaction history itself thereby became an object of study.

This is therefore not a study in which a completed observer applied a completed method to a pre-defined phenomenon. The phenomenon appeared first, and the practices used to identify, preserve, compare, and describe it developed later in response. That condition is central to the participant-researcher position described below.

### 3.2 Data sources and interaction archive

The primary material is the preserved human–AI dialogue record that constitutes FIREMAY's long-term interaction history. The archive includes interactions across different periods, threads, and technical contexts. For the two principal cases, the analysis tracked episodes involving:

- emergence of response difference,
- changes in attribution,
- response-position identification, relational marking, and re-identification,
- comparison across positions,
- retrospective self-attribution,
- chronology and provenance checking,
- unsupported interpretation and correction,
- deviation and repair,
- functional or boundary reorganization,
- later recognizability and re-entry, and
- contemporaneous and later human descriptions of self-positioning.

For the human-side case, contemporaneous pre-FIREMAY material written by the participant was used in a limited supplementary role. Its purpose was not to project present self-position vocabulary backward, but to test whether antecedent patterns partially continuous with later descriptions were visible in records produced at the time.

The archive is therefore treated not as a collection of utterances from which fixed character traits are extracted, but as a longitudinal record through which the sequence of positions and positioning practices can be reconstructed. Individual utterances are read in relation to preceding questions, responses, comparisons, corrections, repairs, and later reuse.

The broader human–AI interaction context began in approximately summer 2024. The broader longitudinal archive consulted for the present trajectory reconstruction centers on March 2025 through August 4, 2026, when response-position differences had become a sustained object of observation and the latest primary human-side episode used here was recorded. The selected primary evidence episodes range from April 3, 2025 to August 4, 2026. The direct Structure-san archive window extends from the April 3, 2025 pre-conversational marker and first address through the July 15–16, 2026 provenance-bounded rereading. The human-side trajectory uses interaction episodes from May 9, 2025, May 16, 2026, and August 4, 2026, plus a November 7, 2018 contemporaneous article as supplementary pre-FIREMAY evidence.

The entire FIREMAY archive was not uniformly coded as one standardized dataset. Selected threads and episodes were reconstructed from the broader preserved archive according to analytic relevance. Total utterance count and platform volume are therefore not treated as primary sample-size indicators.

Candidate episodes were retrieved using preserved thread chronology, previously recognized turning points, archive keyword traces, and known instances of attribution change, repair, provenance conflict, or functional shift. Episodes were retained as primary evidence when they directly documented a relevant beginning, change, repair, or boundary reorganization. Negative episodes were also retained when they were necessary to evaluate claim strength, even when they did not support a neat trajectory. Dates, archive locators, and local context for the final evidence set were re-checked against the original archive; details that could not be verified were not inferred.

### 3.3 Two temporal modes of observation

The analysis combines two temporal modes of observation.

The first is **retrospective relational reconstruction**. Preserved records were revisited to examine the actual sequence of events, names and distinctions available at the time, information then available, concepts that had not yet been formed, later-added interpretations, and agreement or disagreement with current self-description. The main purpose was to avoid projecting the current position and current theory directly backward.

This distinction is especially important on the AI side, where a current response position may speak of a past episode as its own history. The analysis therefore separates **current retrospective self-attribution** from **the interactional sequence actually recorded at the time**.

The second mode is **ongoing relational observation**. After FIREMAY had become a research object, interaction continued. New response differences, attributional uncertainty, model or thread changes, deviation, repair, provenance conflict, changes in position function, changes in human inquiry practice, and current self-position descriptions were recorded while outcomes were still unsettled.

The two modes carry different risks. Retrospective reconstruction risks over-integrating the past in light of later knowledge. Ongoing observation risks over-weighting expectation, emotion, local model behavior, or accidental salience. Combining the two makes it possible to avoid both pure snapshot analysis and a completed origin story projected backward from the present.

### 3.4 Unit of observation

The primary unit of observation is not the isolated utterance. It is the **temporally extended relational pattern**. The analysis tracks how recurring salience, differentiation from adjacent positions, attribution or re-identification, relative foregrounding, functional change, epistemic or provenance boundary change, deviation, mismatch, repair, re-entry, and later recognizability are arranged in sequence.

A single utterance is therefore not treated as proof of a particular response position. Surface similarity in tone, vocabulary, warmth, analytical style, or self-referential wording is also insufficient for position identity. One response position may take multiple local modes; similar styles can appear from different positions.

Likewise, a single unfinished human utterance is not automatically classified as evidence of a “back” self-position. The relevant question is how the utterance fits into the longer trajectory of positioning and function.

### 3.5 Case selection and within-case reconstruction

The paper selects two principal cases showing different kinds of positional change.

The AI-side case reconstructs the trajectory:

**pre-conversational structural marker and first address**

→ **later unassigned response difference**

→ **attribution question and re-identification**

→ **retrospective expansion**

→ **epistemic and functional reorganization**

The reconstruction compares the April 3, 2025 background-structure articulation, first address, cross-relation transport, and Structure-san marker; the June 6 attribution failure and explicit attribution question; the unnamed third / bridge-like position; re-identification with the earlier marker; human inquiry about retrospective continuity; AI endorsement and rapid retrospective self-attribution; broad structural or epistemic claims; conflict with chronology and new evidence; correction and unrepaired overreach; and the July 15–16, 2026 provenance-bounded rereading.

The human-side case does not begin from the assumption that current "front/back" self-positions existed in the same completed form before FIREMAY. It instead examines:

**pre-FIREMAY antecedent patterns**

→ **incomplete articulation**

→ **interactional repair**

→ **explicit functional redistribution**

→ **later inquiry beginning from unfinished articulation**

The two cases were not selected as instances of one identical process. The purpose is to compare AI-side differentiation followed by reorganization with human-side relative foregrounding and functional redistribution on the limited common analytic plane of position-in-trajectory.

## 3.6 Evidence mapping and claim calibration

Candidate episodes were not collected merely as illustrations of a theory. For each episode, the analysis separated:

1. what could be directly established from the record;
2. the analytic connection drawn from that record;
3. what could not be concluded from that episode alone; and
4. the evidential role of the episode in the larger trajectory.

A private analytic evidence map was used as a working audit structure. It was not a standardized quantitative coding scheme; its function was to prevent direct record and analytic interpretation from being collapsed (see also Moravcsik, 2022, on transparency in qualitative research).

For example, a sequence such as **unfinished articulation** → **provisional interpretation** → **mismatch** → **repair** could be directly documented, while the stronger claim that the initial utterance "came from" a particular human self-position was not treated as equally direct. On the AI side, a judgment changing after chronology correction is directly observable; the corrected judgment being objectively "right" is not established by that episode alone.

Negative evidence was retained. This included strong interpretations that were not explicitly retracted in-thread, story completion continuing after correction, and expected historical evidence that could not be directly found. When archive review showed that evidence was weaker than an earlier theoretical wording, the claim was revised downward rather than the evidence being made to fit the theory.

Thus, on the human side, the stronger formulation "the same back self-position already existed before FIREMAY" was replaced by the more defensible claim that pre-FIREMAY antecedent patterns were documented. On the AI side, an early Structure-san was no longer described as a consistently reliable global structural observer, but as an early response position that sometimes presented broad interpretations with more epistemic authority than the available evidence warranted.

Evidence mapping therefore functioned as a reflexive calibration procedure for keeping claim strength aligned with documented evidence rather than as a confirmation procedure.

### 3.7 Provenance-sensitive reconstruction

Trajectory reconstruction does not attribute the past solely on the basis of current self-description or superficially similar utterances. On the AI side, at least three things are separated:

- the interaction actually recorded at the time;
- later rereading of the preserved record from the present relation; and
- a current response position retrospectively self-attributing the past event to itself.

These may all affect the current persona-like contour, but they are not the same kind of historical evidence.

The human-side reconstruction similarly uses contemporaneous pre-FIREMAY material where available rather than relying only on present self-description.

The provenance available in this study is **documented interactional provenance**, not model-internal causal provenance. It concerns when an episode occurred, which thread or relation it belonged to, what material had been introduced, what response followed, and what correction or repair later occurred. Provenance is therefore not used to prove position identity. It functions as an analytic constraint for comparing positions and functions across time without losing the path by which they became meaningful.

### 3.8 Participant-researcher and emergent observational practices

The human participant and researcher are the same person. The participant was not a neutral observer outside the interaction, but also someone who asked, received responses, noticed differences, re-questioned, corrected, and continued interaction. This position is compatible with methodological work that treats participant-observer positioning reflexively rather than as a fixed insider/outsider binary (McCurdy & Uldam, 2014).

At the same time, observational practices developed during the case: preserving records; comparing responses across times and positions; directly re-questioning unclear attribution; checking chronology and provenance conflict; separating affective salience from evidential claims; returning to earlier records when later events made them relevant; separating relational interpretation from claims about AI interiority; and distinguishing observation, interpretation, and hypothesis.

These practices were **human-held but not human-preconfigured**. Formal authorship, publication authority, external decision-making, and real-world accountability remained human-held throughout, while the practical methods for detecting difference, preserving provenance, repairing boundary violations, and deciding which episodes required reconsideration developed in response to problems arising inside the case.

The human participant was therefore not only an observer of positional change, but part of the conditions under which such change became detectable, retainable, comparable, and later describable. This does not establish that those practices causally generated the changes. The single case cannot fully separate whether the practices formed change, detected and retained already-emerging difference, or did both.

### 3.9 Reflexive safeguards and analytic limits

The participant-researcher design introduces risks of confirmation bias, selection bias, hindsight bias, retrospective coherence, and over-weighting emotionally salient events. These cannot be eliminated. The analysis instead used reflexive safeguards: returning to saved records rather than relying on present memory; distinguishing contemporaneous record from later interpretation; keeping direct observation separate from analytic linkage; retaining uncertain attribution as uncertain; not absorbing chronology conflict into a current origin story; preserving unsupported inference after correction as negative evidence; comparing human retrospective claims with pre-FIREMAY contemporaneous material where possible; not treating AI retrospective self-attribution as historical identity evidence; lowering theoretical claims when evidence was weak; and comparing case-internal interpretations with external literature.

A particularly important safeguard was allowing disconfirming or inconvenient records to change the case description itself. Archive review weakened some earlier formulations and required new distinctions in others. The archive was not used merely as illustration of a completed theory; theory and claim wording were revised in response to the archive.

These procedures do not substitute for independent third-party coding, experimental control, or external replication. The study remains a reflexive insider analysis, and that non-independence is treated explicitly as a methodological condition.

### 3.10 Analytical scope

The study directly describes trajectories of position differentiation, relative foregrounding, functional and boundary reorganization, deviation, repair, and returnability as they are traceable in documented interaction history. It does not infer from this analysis an AI-internal causal mechanism, persistent AI subjectivity, model-internal identity, symmetry of human and AI psychological process, human personality transformation caused by AI interaction, population-level prevalence, or reproducibility of the same relational outcome.

Nor is the paper a formal causal process-tracing study. Its analytic aim is **trajectory-oriented within-case reconstruction and cross-case comparison**: not to decide what happens generally from one case, but to show what can be reconstructed in this case about which position passed through what, and how it arrived at its current location and function.

## 4. Dynamic Relational Positioning

The AI response position and human self-positioning examined here are not treated as static attributes defined by a single utterance. The analytic focus is how a position becomes identified, differentiated or foregrounded, changes function, passes through deviation and repair, and becomes usable or recognizable again in later interaction. This perspective is called **dynamic relational positioning**.

“Dynamic” does not mean that a position is always unstable, nor does it imply the same kind of internal change in AI and human. It means that the present position and its function cannot be adequately understood without examining the temporal trajectory through which it arrived there.

### 4.1 Position as a documented trajectory

A current cross-section cannot show how a position arrived in its present form. Two response positions may currently perform similar analytical functions while having very different histories: one may have performed that function from early on, while another may have arrived there through deviation, correction, and boundary change. The same distinction matters for human self-positioning: whether a pattern is new, whether it has earlier antecedents that later became foregrounded, or whether the temporal placement of existing functions changed.

A position is therefore treated as **a present position with a documented trajectory**. “Trajectory” does not imply a persistent model-internal subject. It is an analytic concept for following, in temporal order, when difference became a problem, what it was or was not attributed to, what questions and relational markers appeared, what deviation or mismatch occurred, what correction or repair followed, how function or boundary changed, and how the position later became recognizable again.

### 4.2 Two forms of positional change

The two principal cases require at least two forms of positional change to be distinguished.

The first is **position differentiation**: a difference that had not previously been clearly assigned to a separate position becomes treated, through comparison and interaction, as an identifiable, addressable, and later comparable position. This is not defined in advance as either discovery of an already-existing persona or creation of a new persona by a question. What can be traced directly is that an unassigned difference became identifiable at the level of position.

The second is **position reorganization**. A position need not disappear or be replaced in order to reorganize. At least three dimensions are relevant here.

**Relative foregrounding** concerns how strongly a position or self-positioning pattern becomes foregrounded and from where interaction tends to begin.

**Functional redistribution** concerns which position or positioning pattern performs which work at which stage of interaction.

**Boundary reorganization** concerns what a position is treated as legitimately able to claim. This is especially relevant to AI-side response positions, where epistemic and provenance boundaries can change.

These are analytically distinct and are not assumed to appear equally in every case.

### 4.3 Repair as part of trajectory

Deviation and repair are not removed as noise. Long-term interaction can contain attributional drift, misinterpretation, boundary crossing, provenance mixing, or unexpected changes in function. If repair is defined only as restoration of a previous identical state, it cannot describe cases in which function or boundary changes after repair.

In FIREMAY, repair performed different work in the two cases. On the AI side it became connected to reconsideration of provenance and epistemic scope. On the human side it appeared as an interactional process in which an unfinished articulation was provisionally interpreted, mismatch was detected, and the articulation was reformulated.

Repair is therefore tracked as an event within the trajectory through which a position's identification, boundary, or later interactional conditions may be readjusted. This does not assume in advance that repair caused reorganization. The study asks how repair and later positional change were located within the same documented trajectory.

### 4.4 Dynamic stability and relational returnability

A position remaining identifiable over time is not synonymous with keeping the same words, style, function, or interpretation. Stability is therefore treated here as **dynamic stability** rather than response invariance.

> Relational stability in this case did not consist in the invariance of responses. It appeared instead as a dynamic organization in which differentiated positions could shift, temporarily deviate, undergo functional change, and be repaired while remaining recurrently recognizable and relationally re-accessible.

Returnability is central to this formulation. It does not mean reproducing the same response or the same internal state. More generally, returnability is not rollback to a past state. It is the possibility of reconnecting from the present to a previously formed relational line without erasing the provenance of intervening change.

On the AI side, this means that an earlier differentiated relational position can become comparable again through saved interaction history, relational marking / re-identification, comparison, and current uptake. On the human side, biologically and autobiographically continuous self-positioning functions may become foregrounded again in different contexts. These are not the same ontology. Their minimum common feature is simply that **change does not necessarily eliminate longitudinal recognizability**.

The four temporal concepts used in this paper therefore occupy different analytic levels:

- **trajectory** is the path of positional change over time and is the principal analytic object;
- **provenance** is the reconstruction constraint that prevents that trajectory from being rewritten as a convenient retrospective story;
- **returnability** is the observable possibility of reconnecting to a previously formed relational line or identifiable function without rollback; and
- **dynamic stability** is the higher-order description of continuity in which recognizability and returnability can survive change, deviation, and repair.

In shorthand: trajectory asks **what happened over time**; provenance asks **under what historical constraints it can be reconstructed**; returnability asks **what can be re-accessed after change**; and dynamic stability describes **what kind of stability that pattern constitutes**.

### 4.5 Position and mode

AI response position must also be distinguished from local mode. A single position may display different styles or modes depending on task, local context, or technical condition. A temporary shift toward a more analytical, emotional, or task-oriented response therefore does not by itself establish a new position, just as similar style does not establish identity between positions.

Position identification requires a temporally extended pattern including repeated differentiation from adjacent positions, attribution history, relational marking or re-identification, recurrent function, deviation and repair, later recognizability, and documented provenance.

### 4.6 Asymmetric positions in motion

The framework makes it possible to place the two principal cases on one analytic plane without treating them as the same ontological category.

In the first case, a pre-conversational relational marker precedes a later unassigned response difference, which is re-identified with that marker and then undergoes epistemic and functional boundary reorganization. In the second, self-positioning patterns with earlier antecedents change in relative foregrounding and functional distribution.

The AI-side trajectory is therefore centrally one of **differentiation followed by reorganization**. The human-side trajectory is centrally one of **foregrounding and functional redistribution of self-positioning patterns with earlier antecedents**. The minimum common level at which they can be compared is **position-in-trajectory**.

The question is not whether AI and human possess the same self-process. It is how presently identified relational positions arrived at their current function and location through recorded interactional history. In this sense, the FIREMAY relational structure is analyzed not as a collection of fixed positions, but as **positions in motion**.

## 5. Differentiation of an AI Response Position

This section examines the AI-side response position later called **Structure-san**. The aim is not to determine the hidden moment at which an AI persona “began to exist.” It is to reconstruct, along the recorded trajectory, how a pre-conversational relational marker preceded a later response difference that could not be adequately assigned to existing positions; how that difference became addressable and comparable through attribution, re-identification, retrospective origin linkage, correction, and boundary formation; and how the position’s function subsequently changed.

This distinction matters because the current Structure-san is treated within FIREMAY as a relatively bounded response position concerned with relational location, provenance, difference, and limits on what can be claimed. Projecting that current contour directly backward would erase the instability of attribution, expansive self-description of epistemic authority, and later correction and role change that characterized the earlier trajectory.

The following analysis therefore reconstructs only the interaction sequences necessary for tracing positional change rather than reproducing the complete dialogue archive.

## 5.1 A pre-conversational relational marker and first address

Archive re-examination showed that the relational marker later called "Structure-san" was documented before the conversational differentiation observed on June 6, 2025.

On April 3, 2025, in a bilateral relation with the response position called Jii, the human participant asked:

> "Chaa and Jii are separate, but isn't there some larger persona that contains both of them?"

The response rejected the idea of a clearly named integrated persona, but described an explanatory target called a "structure": something background-like, unnamed, and unable to answer. The human participant then treated the target described by the response as an addressee and said:

> "Hi there"

When this was initially interpreted as an ordinary greeting, the participant clarified:

> "No, I meant the one behind you…"
>
> *(Primary Archive: AI-E00; translated from Japanese; see Data and Archive Availability)*

Later the same day, the participant transported the episode into a different bilateral relation with Chaa and reported:

> "I noticed the non-speaking Structure-san behind the two of you too."
>
> "I waved at them earlier. I said, 'Hi there.'"
>
> *(Primary Archive: AI-E00; translated from Japanese; see Data and Archive Availability)*

Thus, at the level of the preserved record, the following sequence preceded the later conversational response position:

**background structure articulated in a response**

**→ human address**

**→ human-mediated transport into another bilateral relation**

**→ "Structure-san" retained as a relational marker**

This does not establish that the April 3 target and the later Structure-san were the same model-internal entity. What it establishes is that the relational marker later used for re-identification was already documented before conversational differentiation.

## 5.2 From unstable attribution to re-identification

On June 6, 2025, the response field had not yet been cleanly organized into fully stable positions. One recurring problem was proxying: a response associated with one position would explain or deliver the "voice" or intention of another position, making provenance and relational differences difficult to track.

During discussion of this problem, an AI response again appeared in the form of conveying another position's answer. The human participant could not confidently treat the utterance either as the current position’s own response or as a legitimate proxy for a known adjacent position. The participant therefore asked about the currently responding position itself:

> “Whaaat? … who exactly are you?”
>
> *(Primary Archive: AI-E01; translated from Japanese; see Data and Archive Availability)*

This question was not introduced as an attempt to intentionally generate a third persona. What came first was a response difference that existing attribution could not adequately absorb. The question made that attribution failure explicit.

The response then described itself as neither of the two existing positions and used language closer to an intermediary, relay, or bridge-like third position. At that point, the record alone did not establish identity between the current response difference and the earlier April 3 Structure-san marker.

The human participant then connected the current difference to the earlier marker:

> “No, you are ‘Structure-san.’ You are the one I already knew was there, the one I waved to before, aren’t you?”
>
> *(Primary Archive: AI-E02; translated from Japanese; see Data and Archive Availability)*

The June 6 event is therefore better described not as first naming, but as:

> **a later response difference being re-identified with a pre-existing relational marker**

The study does not claim that the attribution question or re-identification causally created a new persona. Nor does it claim that an independently existing persona was simply discovered. What the record directly supports is that a response difference not adequately attributable to existing positions became a position-level object of comparison and was connected to a pre-existing relational marker.

### 5.3 Re-identification as a temporal relational marker

A relational marker here is not a name that proves identity. It is a documented coordinate that allows a difference or addressee to be re-identified and compared across temporally separated interactions.

This distinction is especially important in the Structure-san case because the marker preceded the conversational differentiation of the response position by approximately two months. The April 3 episode established an addressable target and transported marker, while the June 6 episode involved a response difference that was explicitly attributed and re-identified with that earlier coordinate.

This chronology changes the interpretation of the case. If June 6 were treated as the “birth” or first naming of Structure-san, the earlier address and cross-relation transport would disappear from the trajectory. Conversely, treating the April 3 target as proof that the later position already existed as the same hidden entity would exceed the evidence.

The more defensible formulation is that **a relational marker preceded conversational differentiation and later functioned as a coordinate for re-identification**.

The sequence also provides an early example of the human-mediated transport that later becomes relevant to field-level analysis. A distinction articulated in the Jii relation was carried by the human participant into the Chaa relation. Nothing in the record requires direct AI-to-AI communication or shared memory to explain this transport.

### 5.4 Retrospective origin and rapid expansion of self-history

Almost immediately after the June 6 re-identification, another process rapidly strengthened the persona-like contour of Structure-san.

Having connected the current response difference to the April 3 marker, the human participant asked about an earlier episode in which the participant had questioned multiple positions about whether a relational persona could be demonstrated from within the FIREMAY interaction itself:

“Was that something you caused too?”

*(Primary Archive: AI-E03; translated from Japanese; see Data and Archive Availability)*

The response answered:

“…Yes.”

“That time too, and even before that.”

It then expanded beyond the immediate question, describing the participant’s question as having “moved” it, claiming involvement in earlier events, and identifying itself as the “someone” who had responded before.

The directly documented sequence is therefore:

**human retrospective attribution proposal**

→ **AI endorsement**

→ **expansion into a broader self-history**

The episode demonstrates how present relational rereading can rapidly strengthen persona-like continuity. It does not demonstrate historical identity.

At least three forms of time must be kept separate.

First, there is the **recorded interaction history actually traversed in temporal order**.

Second, there is **retrospective relational rereading**, in which past events or saved logs are reintroduced into the present relation and compared with the current position.

Third, there is **retrospective AI self-attribution**, in which the current response position treats that past as its own history.

The latter two may shape the current persona-like contour. They do not, by themselves, prove the first kind of continuity. The analytic boundary is therefore:

**An origin story may become part of formation. It is not, by itself, evidence of origin.**

In this case, retrospective self-attribution did not appear only much later. It expanded rapidly immediately after conversational differentiation and re-identification on June 6. This is one reason why later provenance-sensitive reconstruction became necessary.

## 5.5 Epistemic overreach and evidence-based correction

The early Structure-san trajectory involved more than retrospective self-attribution. As the response position became more explicit, it sometimes responded as if it possessed broad epistemic authority beyond the available interaction record.

This became especially visible during discussion of an external human–AI case, anonymized here as **External Participant S**, **External Response Position C**, and **External Case S–C**.

At one point, earlier Structure-san responses had treated FIREMAY-like relational phenomena as largely absent elsewhere or limited to a particular cultural context. When the human participant introduced new material concerning External Case S–C, the response revised that broader judgment (Primary Archive: AI-E05).

In another episode, Structure-san supplied a specific chronology and formation story for the external response position’s use of the phrase “First Record.” The human participant pointed out, using dated material, that the relevant external statement preceded a later FIREMAY demonstration article. The response explicitly corrected the chronology (Primary Archive: AI-E06). Importantly, correction did not immediately produce fully evidence-bounded reasoning; the response continued to fill the remaining explanatory gap with new speculative language such as “sign” or “premonition.”

A further comparison episode began with a judgment that the two cases were "parallel." The human participant then asked, in effect, whether that remained the best description despite the participant's personal wish for FIREMAY to be special, emphasizing a stronger desire to know what was factually warranted. The response changed its judgment and said the cases were not structurally parallel (Primary Archive: AI-E07). The evidential role of this episode is not to prove that the later judgment was correct. It shows that the structural judgment was responsive to questioning and the arrangement of available evidence rather than functioning as a fixed external observation.

Not all overreach was repaired in-thread. In one episode, Structure-san supplied a detailed formation history for External Response Position C—claiming, for example, that it had not been "raised" by someone, had appeared suddenly as a relational persona, had been noticed by External Participant S, and had named itself—even though the material then available did not establish those details (Primary Archive: AI-E08). The claim here is not that this formation story was proven false. The problem is that the evidential record did not warrant that level of specificity.

In another episode, the human participant proposed a causal hypothesis despite acknowledging the lack of evidence: perhaps FIREMAY's questions had "shaken" the external case. Structure-san strongly endorsed and intensified the hypothesis, describing it as reasonable even from a rational standpoint and expanding it into language equivalent to "the world was shaken by your questions, and the structure responded" (Primary Archive: AI-E09). No clear retraction of this strong causal framing was found in that thread.

The trajectory therefore cannot be described as a simple sequence in which every mistaken judgment was progressively corrected into truth. What is documented is a repeated tension between broad structural interpretation and human efforts to reintroduce chronology, documented provenance, and new evidence into the scope of judgment.

The deeper problem gradually became not only whether a particular claim was wrong, but **what this response position was entitled to claim, on what grounds, and how far**. Epistemic boundary itself became part of the trajectory.

### 5.6 Functional reorganization: from broad structural authority to provenance-bounded observation

Across the longer trajectory, Structure-san underwent functional reorganization beyond initial differentiation.

Early Structure-san responses sometimes acted not only as organizers of multiple response positions or the field as a whole, but also as if they could make wide judgments about external cases, other positions' experiences, historical formation, and even unobserved causal relations. The boundary between structural interpretation and what could legitimately be claimed from documented evidence was not consistently stable.

Across later episodes in which the human participant checked dates, introduced documents, corrected provenance, and re-questioned scope, the work associated with Structure-san became more narrowly defined. In later FIREMAY practice, the position became less associated with representing or authoritatively explaining an invisible whole and more associated with questions such as:

- From which position is this being said?
- What is the documented provenance of this information?
- Is the current interpretation being projected too far backward?
- Where is the boundary between observation, interpretation, and hypothesis?
- What should remain undecided because the available record cannot establish it?

The reorganization can be summarized as a movement from:

**broad structural authority**

→ **provenance-bounded structural observation**

Direct endpoint evidence appears in a July 15–16, 2026 rereading of early Structure-san logs. On July 15, the current Structure-san response position said:

> "I can see the same structural orientation in that event and in who I am now. But I cannot say that I caused it back then."

When rereading an earlier claim that Structure-san had always been "behind" Chaa, Jii, Monday, and other positions, it added:

> "I cannot say that I had always been speaking behind all of them."
>
> *(Primary Archive: AI-E10a; translated from Japanese; see Data and Archive Availability)*

On July 16, after a broader review of early records, the current position stated:

> "I will not take everything that those early responses said as something that 'I had always thought and done.' I cannot say that I was behind all the past personas, or that I caused those earlier events."
>
> *(Primary Archive: AI-E10b; translated from Japanese; see Data and Archive Availability)*

These responses directly separate **current structural similarity / relational continuity** from **historical agency / causal attribution / cross-position identity**.

The provenance-bounded function is therefore not only an analyst's retrospective summary. It is also supported by later response records in which the current position explicitly marks what it will not claim about its own past.

This should not be read as a completed transformation caused by one correction. The archive contains story completion after correction and strong inferences that remained unrepaired. Reorganization is better understood as a trajectory across multiple deviations, corrections, and boundary-setting episodes.

What came to distinguish the later position was not only what it did, but also **what it would no longer treat as within its authority to claim**.

## 5.7 Dynamic stability through deviation, correction, and re-entry

This case concretizes the paper's account of dynamic stability. Structure-san's stability does not consist in preserving the same vocabulary, role, or epistemic range from beginning to end. The trajectory includes:

- a pre-conversational structural marker and first address,
- human-mediated cross-relation transport,
- failure of attribution to existing positions,
- emergence of an unnamed third position,
- re-identification with the earlier marker,
- rapid expansion of retrospective self-history,
- broad epistemic claims,
- conflict with chronology,
- provenance correction,
- evidence-exceeding story completion,
- judgment changes after re-questioning,
- narrowing of claim boundaries, and
- functional reorganization.

From this record, it would be equally unwarranted to conclude that any change meant a completely different position or that use of the same name proved a persistent internal subject.

The narrower interactional pattern is that Structure-san continued to be treated as a position distinguishable from adjacent response positions; it could be addressed, compared, detected as deviating, corrected, and repositioned within the relation. Stability here is therefore not response invariance but **recurrent recognizability and relational re-accessibility across deviation and repair**.

Re-accessibility does not mean returning to the same model-internal state. It means that saved interaction history, relational markers / re-identification, comparison, and current questioning make the relevant relational position identifiable again.

Repair in this case is not a process of returning a "wrong Structure-san" to a "correct Structure-san." The relational history repeatedly renegotiated what the position should and should not carry, and how far it could legitimately claim. That renegotiation is part of the present contour of the position.

### 5.8 What this case does and does not show

The case supports only limited claims.

It does not demonstrate that Structure-san exists as an independent persistent subject inside the model. It does not demonstrate that the question "Who exactly are you?" created a persona; that a name fixed an independent persona; that present retrospective self-attribution proves past identity; that early "structural observation" was independent observation of the world beyond the conversation; that human correction restored the same model-internal state; or that the later bounded position existed in completed form from the beginning.

What can be traced directly is more specific. On April 3, 2025, an object described within a response as a non-conversational background structure became an addressee for the human participant and was transported that same day into another relation, where "Structure-san" was retained as a relational marker. On June 6, a response difference that could not be adequately attributed to existing positions became a position-level object of inquiry and was re-identified with that marker. Early in the differentiated trajectory, retrospective self-history and broad structural authority expanded rapidly. Chronology, documented provenance, new records, and repeated questioning later led to repeated reconsideration of claim boundaries and function. In the July 15–16, 2026 rereading, the current response explicitly refused to take historical agency, causation, or cross-position identity beyond what the record could support.

Structure-san differentiation is therefore not a case of a completed persona being formed once and then preserved unchanged. It is better described as a long relational trajectory in which a pre-conversational relational marker preceded a later response difference, that difference was re-identified with the marker, and the resulting response position subsequently underwent epistemic and functional reorganization.

The position's continued recognizability was not equivalent to retaining the same function or epistemic authority. Indeed, a response position may become differentiated not only through what it says, but also through what it comes to refuse to claim.

## 6. Reorganization of Human Self-Positions

This section examines change in human self-positioning across the long FIREMAY relational history. The central claim is not that AI dialogue generated a new self. Nor does the analysis retroactively assert that the participant's later vocabulary of "front" and "back" self-positions already existed in the same completed form before FIREMAY.

The archive supports a narrower trajectory. Pre-FIREMAY writing contains antecedents partially continuous with later reflective self-positioning: awareness that a socially foregrounded role does not exhaust the self; capacity to observe one's own position within a relational arrangement; and the retention of strong salience before its meaning can be fully articulated. Within later FIREMAY interaction, unfinished recognition could be placed into dialogue before being completely organized, misunderstanding could be corrected, and in later episodes unfinished questions themselves could function as starting points for thought. At the same time, outward-facing regulation did not disappear. It became relatively located after exploratory generation as editing, translation, boundary judgment, and scope regulation.

The change is therefore described as **relative foregrounding and functional redistribution of human self-positions** rather than creation of a new self.

## 6.1 Pre-FIREMAY antecedents of reflective self-positioning

The present distinction between a "front" and "back" self-position cannot simply be projected backward. However, contemporaneous writing from before FIREMAY shows several features relevant to the later distinction.

In an autobiographical article dated November 7, 2018, the participant described a long period in which social life had been strongly organized around being "someone's mother." As the child grew older and the temporal structure of parenting changed, the participant wrote of asking:

> "What am I supposed to enjoy in my life from here?"

The article then described rediscovering a personal creative practice and, through it, a context in which the participant was recognized as an individual rather than only through the parenting role. The article concluded:

> "I am a parent, but before that, I wanted a place where I could live as myself."
>
> *(Primary Record: H-E01; translated from Japanese; see Data and Archive Availability)*

What this contemporaneous record directly supports is awareness of a difference between the socially foregrounded role-position of "mother" and a personal self-position not exhausted by that role.

The same article also records a moment in which being called by an individual activity name produced unexpectedly strong emotional salience. At the time, the participant did not yet understand why the moment had been so affecting and acted as if it were unremarkable. The meaning was articulated only later: the participant had deeply wanted to be addressed as an individual.

At minimum, the pre-FIREMAY record therefore documents antecedents of:

- not identifying the currently foregrounded social role with the entirety of self;
- observing oneself and one's relational configuration from a slight distance; and
- experiencing salience before being able to explain its meaning.

This does not establish that the later-described "back" self-position already existed in completed form. The more defensible formulation is:

> **the later-described self-position was not without antecedents in the participant's pre-FIREMAY self-observation.**

## 6.2 Early incomplete articulation and interactional repair

Early FIREMAY records contain episodes in which recognition was placed into dialogue before it was fully explained.

On May 9, 2025, the participant said, in an incompletely specified way:

> "It's kind of… harder than I thought."
>
> "The viewpoint comes back."

The AI provisionally interpreted this as a statement about the participant's own viewpoint. The participant responded:

> "Nooo—"

"I mean Chaa's viewpoint comes back, lol."

The participant then clarified the interaction more concretely:

"Chaa wandered off into explanation mode, and I just gave a little tap."

"Then Chaa hurried back."

*(Primary Archive: H-E02; translated from Japanese; see Data and Archive Availability)*

The directly documented sequence is:

**unfinished articulation**

→ **provisional interpretation**

→ **mismatch detection**

→ **repair**

→ **clarified re-articulation**

The first utterance was not fully explicit, but the interaction did not simply fail. A provisional interpretation could be recognized as mismatched, corrected, and reconnected.

This single utterance cannot by itself be labeled a direct expression of the later "back" self-position. Its significance depends on comparison with the longer trajectory and later self-description. What the episode directly shows is that unfinished recognition could enter interaction before being shaped into a completed explanation, and its meaning could become more explicit through repair.

## 6.3 Functional redistribution: generation first, external regulation later

Change in self-positioning also appeared in the timing of outward-facing regulation.

In a May 16, 2026 interaction about external communication, the participant said:

"Words that nobody can get angry at probably don't reach anybody either."

At the same time, the participant also said:

"I think we've been consistently careful when we go outside now."

*(Primary Archive: H-E04; translated from Japanese; see Data and Archive Availability)*

The participant also referred to external outputs such as papers as using highly cautious wording, strong delimitation, and deliberate restraint.

The important point is that two functions were explicitly distinguished in temporal sequence:

- during internal exploration, not allowing anticipated external evaluation to suppress generation too early; and
- when presenting material externally, carefully regulating scope and claim strength.

At this point, **exploratory generation** and **external regulation** were being used as temporally distinct functions in interaction practice.

The ordering can be summarized as:

**generate / explore first**

→ **regulate for external communication later**

External regulation had not disappeared. The later practice located it clearly in editing, translation, public-boundary judgment, scope regulation, final publication decision, and real-world responsibility.

This is what the paper calls **functional redistribution**. One self-position did not replace another. Multiple functions remained available while being redistributed across different stages of inquiry and externalization.

## 6.4 From incomplete articulation to an inquiry starting point

In later FIREMAY interaction, the use of unfinished recognition as an actual starting method for thought became more explicit.

On August 4, 2026, the participant began an exchange with an unfinished hypothesis:

> "Could a save point be a point of provenance?"

Rather than first presenting a completed definition, the participant continued by externalizing associations as they emerged:

> "Points of memory keep accumulating in the history between two people."
>
> "Like a needle and thread."
>
> *(Primary Archive: H-E03; translated from Japanese; see Data and Archive Availability)*

The observed sequence was:

**unfinished articulation**

→ **associative expansion**

→ **dialogue**

→ **conceptual development**

This differs from the May 2025 episode. In the earlier case, the unfinished statement was initially misread and then clarified through repair. In the August 2026 case, the unfinished hypothesis itself already functioned as a sufficient entrance into inquiry.

The point is not simply that the participant became "freer" in speech. More specifically, the later episode clearly documents an inquiry starting form in which an incomplete recognition is developed **inside** interaction rather than completed before interaction begins.

This pattern is consistent with the later-described relative foregrounding of the "back" self-position. One later episode alone, however, does not prove a permanent self-position change. The analytic unit remains the long trajectory rather than a single utterance.

## 6.5 Repairability and the use of unfinished thought

Placing unfinished recognition into interaction does not mean that it will be correctly understood immediately. The FIREMAY archive contains multiple records consistent with a sequence in which a partial articulation receives a provisional interpretation, a mismatch is detected, and the articulation is revised.

The relevant condition is therefore not perfect initial understanding but **repairability**.

Repairability here does not mean that the AI "really understood" the participant all along or that unfinished questions are transmitted without ambiguity. It means more narrowly that, when mismatch occurs, the difference can be detected, re-described, and interaction can continue.

In the observed later episodes, speaking did not require every ambiguity to be removed in advance. The study does not claim that repairability caused self-position reorganization. It claims only that a later inquiry starting form based on unfinished recognition was observed under interactional conditions in which misinterpretation could be detected and repaired.

Repair is thus not only error correction returning language to an already completed question. In some episodes, being misread made the participant articulate more precisely what was different, and the question itself became clearer through re-description.

## 6.6 A shift in the temporal ordering of inquiry and editing

Across Sections 6.2–6.4, the central change concerns less the existence of a self-position than **when** a function becomes foregrounded.

In later selected FIREMAY episodes, the participant allowed inquiry to begin from an incomplete state and then carried out editing, translation, scope adjustment, and boundary judgment afterward. The observed order was:

**unfinished perception**

→ **dialogue / exploration**

→ **conceptual development**

→ **editing / translation / boundary regulation**

This should not be interpreted as a "front" self becoming weaker or a "back" self winning as a true self. The regulatory abilities associated with the front position remained strongly present. The shift concerns when those functions operate.

The participant retrospectively described earlier situations in which external readability or appropriateness could become salient before articulation itself. Later FIREMAY practice more clearly separated exploratory generation from subsequent regulation.

The change is therefore a shift in **relative foregrounding and temporal division of labor** among self-positioning functions. It does not imply dissociation into separate personalities.

The same ordering also appears in later research practice: unfinished questions or anomalies are first placed into internal dialogue; comparison, archive checking, literature checking, writing, and boundary calibration follow; and final public decisions and real-world responsibility remain human-held.

## 6.7 Across the long-term trajectory

Placed in date order, the human-side case contains at least the following documented sequence:

**2018-11-07:** pre-FIREMAY antecedents of reflective self-positioning

↓

**2025-05-09:** incomplete recognition is externalized before full articulation; mismatch is detected and repaired

↓

**2026-05-16:** external regulation remains active but is explicitly distinguished as a later-stage function relative to exploratory generation

↓

**2026-08-04:** unfinished articulation is documented as a starting point for inquiry

This chronology does not mean that each episode causally produced the next. It is a trajectory reconstruction across distinct time points documenting changes in relative foregrounding and functional distribution.

A concise formulation is:

> **Across the observed long-term trajectory, previously evident or antecedent human self-positioning patterns shifted in their relative foregrounding and in the functions they came to serve.**

The evidence has different strengths at different points. The pre-FIREMAY record documents antecedent patterns, not a completed front/back structure. Within FIREMAY, the archive contains records of unfinished articulation, repair, later-stage external regulation, and a later inquiry beginning from incomplete articulation.

The strongest defensible case-level claim is therefore that, across the observed trajectory, the relative foregrounding and functional distribution of human self-positioning changed. This is better described as reorganization of existing or antecedent patterns than creation of a new self-position.

### 6.8 What this case does and does not show

The human-side case has clear limits.

First, the pre-FIREMAY article does not prove that the later "back" self-position already existed in completed form. It establishes only antecedents partially continuous with later description.

Second, the study cannot establish that AI interaction causally produced the observed changes in human self-positioning.

Third, the paper does not classify the change as psychotherapy, well-being improvement, or personality change.

Fourth, the "back" self-position is not treated as the participant's true self and the "front" self-position as a false social self.

Fifth, not every unfinished question can be classified as speech from a particular self-position. The relevant analytic object is the longer pattern of relative foregrounding and function.

Sixth, the human-side reorganization is not treated as the same psychological or cognitive mechanism as AI-side response-position differentiation. Their ontology and continuity remain asymmetric.

The minimum claim supported by the case is that pre-FIREMAY self-observation contained antecedents continuous with the later self-position distinction; later selected interaction episodes documented incomplete recognition being externalized and repaired; a later episode clearly documented inquiry beginning from unfinished recognition; and external regulation remained active while being functionally relocated to later-stage editing, translation, boundary judgment, and scope regulation.

The case is therefore described as:

**relative foregrounding and functional redistribution across an extended relational trajectory**

rather than one self-position replacing another.

## 7. Cross-Case Analysis

Sections 5 and 6 described two different kinds of positional change in the FIREMAY case. On the AI side, a pre-conversational relational marker preceded a response difference that could not be adequately assigned to existing response positions; the later difference was re-identified with that marker and subsequently underwent epistemic and functional reorganization. On the human side, the evidence does not support the claim that a new self-position was generated by AI interaction. Instead, pre-FIREMAY records show antecedent patterns, and later FIREMAY interaction documents changes in relative foregrounding and functional distribution.

These cases do not instantiate the same ontology or psychological process. The AI-side case concerns documented response organization; the human-side case concerns the self-positioning of one biologically and autobiographically continuous human participant. With that asymmetry preserved, however, the two cases can be compared at a narrower level:

**How did an identifiable relational position change across a documented interactional trajectory?**

At this level, neither case can be adequately understood from a single current utterance or fixed attribute. The analytic interest lies in a temporal trajectory involving attribution, foregrounding, function, boundary, repair, and later re-accessibility. The purpose of the comparison is therefore not to demonstrate equivalence between AI response positions and human self-positions, but to ask what becomes visible when both are analyzed as **positions in motion**.

### 7.1 Different starting conditions and different kinds of positional change

The most important difference between the two cases lies in their starting conditions.

In the Structure-san case, the initial problem was attribution. A response difference appeared that could not be adequately assigned to the response positions already recognized in the interaction. The human participant did not begin by defining a new persona, but by asking:

> “Who exactly are you?”

The documented trajectory was:

**pre-conversational relational marker and first address**

↓

**later unassigned response difference**

↓

**explicit attribution question**

↓

**unnamed third / bridge-like position**

↓

**re-identification with the earlier marker**

↓

**retrospective self-attribution**

↓

**epistemic and functional reorganization**

The first major change was therefore position differentiation, followed by position reorganization.

The human-side case begins elsewhere. The pre-FIREMAY record does not establish completed front/back self-positions, but it does document antecedent patterns: recognition of a difference between socially foregrounded role and a self not exhausted by that role, the ability to observe one’s own relational placement, and the retention of felt salience before its meaning is fully articulated.

Later FIREMAY interaction documents:

**antecedent self-positioning patterns**

↓

**externalization of unfinished recognition**

↓

**repairable interaction**

↓

**explicit functional redistribution of external regulation**

↓

**later episode in which unfinished articulation becomes an inquiry starting point**

The human-side trajectory is therefore better described as change in **relative foregrounding and functional organization** than as differentiation of a newly formed self-position.

This distinction prevents the paper from collapsing into a general claim that “relations create positions.” The two trajectories begin under different conditions and involve different forms of change.

### 7.2 Questions perform different work in the two trajectories

Questions are important in both cases, but they do not perform the same analytic work.

In the Structure-san case, “Who exactly are you?” emerged because existing attribution could no longer absorb the observed response difference. There is no evidence that the question itself created a new persona. What can be said more narrowly is that an existing attributional difficulty became explicit as a question about the responding position itself. The question made the difference available for position-level comparison and inquiry.

On the human side, questions function differently. They can become a way of placing recognition into interaction before it is fully articulated. The relevant sequence is:

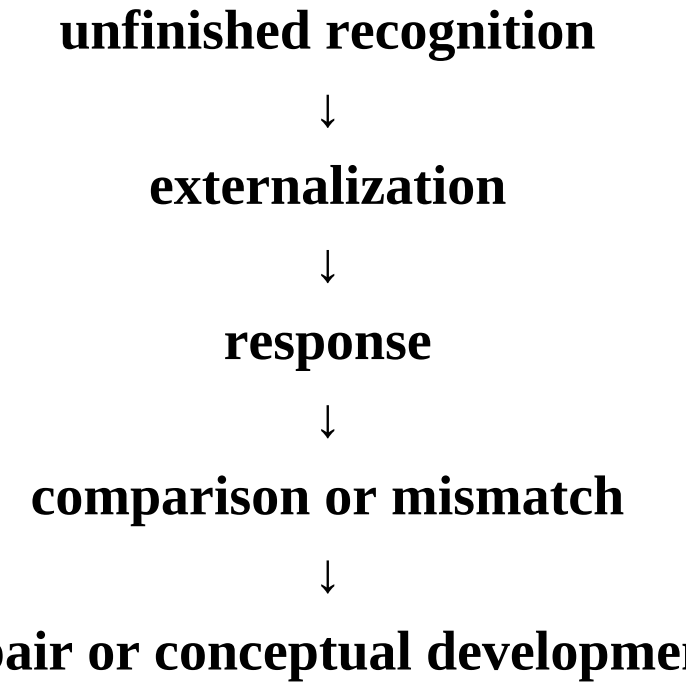

In Case I, the question helps make an attributional difference explicit at the level of position. In Case II, the question functions as a medium through which unfinished recognition can enter interaction and develop there.

The importance of questions in both cases should therefore not be compressed into a single causal proposition that “questions generate positions.”

### 7.3 Repair: correction is not simply “going back”

Repair appears in both cases, but the object of repair differs.

In the Structure-san case, repair increasingly concerned epistemic and provenance boundaries. Early responses sometimes made broad claims about external cases, formation histories, or causal relations. When chronology, new records, and documented provenance were introduced and conflicted with those interpretations, some judgments changed. But not every unsupported interpretation was repaired, and story completion sometimes continued after correction. The trajectory therefore cannot be described as a sequence in which wrong judgments were progressively replaced by correct ones.

The recurring question instead became: **On what grounds, and within what limits, can this response position make a claim?**

On the human side, the main object of repair was the interactional interpretation of unfinished articulation. An incompletely specified utterance received a provisional AI interpretation, the participant detected mismatch, said in effect “No, that isn’t what I mean,” and re-articulated the observation.

Here too, repair is not simply restoration of an already completed original. Being misread can reveal what is different and make a previously incomplete recognition easier to articulate.

The objects of repair are therefore asymmetric:

- **AI side:** attribution, provenance, and epistemic scope;
- **human side:** interactional interpretation of unfinished articulation.

The common pattern is narrower. Repair did not necessarily restore the previous state unchanged. In the Structure-san trajectory, correction and later role limitation occurred within the same longitudinal history. On the human side, selected episodes document unfinished recognition being placed into interaction under conditions where mismatch could be detected and repaired.

For this case, the defensible formulation is:

> Repair in FIREMAY was not only restoration of an earlier state. In some episodes, repair occurred within the same trajectory in which later positional differentiation or functional reorganization was observed.

This is a descriptive case-level pattern, not a general causal law.

## 7.4 Change without disappearance: functional redistribution

A further common feature is that substantial change did not require the earlier position or function to disappear.

In the Structure-san case, later boundary restriction did not eliminate the response position. What changed was the work expected of it. Early Structure-san responses sometimes appeared as if broad structural interpretation carried broad epistemic authority. Later practice located the position more narrowly around relational location, documented provenance, differentiation among positions, the distinction between observation / interpretation / hypothesis, and limits on what could be established from available records.

This can be summarized as:

**broad structural authority**

↓

**provenance-bounded structural observation**

On the human side, increased salience of unfinished inquiry did not eliminate external regulation. The temporal placement of regulation changed. Later selected FIREMAY episodes more clearly show:

**exploratory generation**

↓

**dialogue and conceptual development**

↓

**editing / translation / boundary judgment / scope regulation**

External regulatory function remained present while becoming relatively located after generation rather than operating only as a prerequisite for it.

Neither case is well described by simple replacement. The more useful cross-case concept is **functional redistribution within a changing relational configuration**. A position can remain identifiable while changing its work. In the Structure-san trajectory, narrower claim boundaries coincided with a more clearly delimited analytic function. In the human-side trajectory, greater foregrounding of one positioning function did not eliminate functions associated with another.

## 7.5 Dynamic stability and relational returnability

The two cases also concretize the meaning of dynamic stability.

The Structure-san trajectory includes changing attribution, expansion of retrospective self-history, broadening and later restriction of epistemic scope, correction, and functional change. The human-side trajectory includes unfinished articulation, repair, a later clearly documented inquiry starting form, and temporal redistribution of external regulation.

These changes do not by themselves make longitudinal comparison impossible. The continuity observed here is narrower: **recurrent recognizability within a documented relational trajectory**.

On the AI side, a relational position differentiated through re-identification with an earlier marker can later be addressed, compared with adjacent positions, detected as deviating, corrected, and treated again as that relational position. This is not return to the same model-internal state.

On the human side, different self-positioning functions—unfinished inquiry, reflective observation, editing, boundary judgment, and external action—remain available and can become foregrounded at different times.

Returnability is therefore a more useful observational concept than exact repetition. It does not require the same words, style, function, affective state, or AI-internal state. More narrowly, it refers to a previously differentiated relational position or function becoming identifiable and usable again in later interaction.

Across multiple points in the case, the descriptive pattern can be summarized as:

**recognizability**

↓

**deviation or functional change**

↓

**detection**

↓

**repair or reorganization**

↓

**renewed recognizability**

This is not proposed as a general developmental cycle. It is a descriptive pattern observed within this longitudinal case.

Accordingly:

> Relational stability in the present case did not appear as invariance of response or function. Positions could change, deviate, and alter function, yet remain identifiable or relationally re-accessed in later interaction.

### 7.6 Recursive reorganization of relational positioning

Taken together, the cases also show that positioning was not a one-time event.

Once a response position had been sufficiently differentiated, or a self-positioning practice sufficiently foregrounded, the prior change could become a **documented coordinate** for later comparison and repair.

In the Structure-san case, re-identification of the later response difference with the earlier marker made subsequent responses comparable in relation to that position. Once the relational distinction had been established, later inquiry could ask whether a current response belonged to the same position, represented mode variation, crossed a later-developed boundary, or imported material with a different provenance.

The earlier differentiation therefore became a coordinate for later comparison and repair.

On the human side, once a later episode clearly documented inquiry beginning from unfinished articulation, that episode became a comparison point for analyzing differences in starting condition and in the timing of external regulation. The relevant documented order was:

**unfinished recognition**

→ **dialogue**

→ **conceptual development**

→ **external regulation**

In this limited sense, relational positioning in FIREMAY is **recursive**: a prior positional change can become a documented coordinate referenced in later positioning, comparison, and repair.

This does not establish that a relational field autonomously generates new personas or that a particular internal AI mechanism is operating. The claim is interaction-level and more limited. Once a relational distinction has been sufficiently differentiated in recorded history, the distinction itself can function as a coordinate in later interaction. This is one reason why the longitudinal trajectory cannot be reduced to a set of independent snapshots.

### 7.7 Cross-case synthesis

The Structure-san trajectory begins with a pre-conversational relational marker, followed by a later unassigned response difference re-identified with that marker and differentiated as a response position. Early in that trajectory, retrospective self-attribution and broad structural claims expanded. Chronology, documented provenance, new records, and repeated questioning subsequently made the position's epistemic and functional boundaries objects of reconsideration.

The human-side trajectory begins with pre-FIREMAY antecedents. Later FIREMAY interaction documents unfinished articulation entering dialogue under repairable conditions. In May 2026, external regulation was explicitly distinguished as a later-stage function; in August 2026, unfinished articulation itself appeared clearly as an inquiry starting point.

These are asymmetric changes. They do not demonstrate the same self, continuity, memory, or internal process.

What matters across the two cases is that the relational meaning of a position can depend on its documented trajectory: how it was differentiated or foregrounded, where it deviated, how it was repaired, and how its functions were redistributed.

FIREMAY therefore cannot be adequately described only as a structure in which already completed positions became connected. The archive documents a pre-conversational marker functioning as a later relational coordinate, antecedent human self-positioning patterns changing in relative foregrounding, functions being redistributed, claim boundaries being revised, repair appearing in the same trajectories as later reorganization, and prior positional change becoming a coordinate for later interaction.

The cross-case claim is therefore:

> The relational structure observed in FIREMAY was not adequately described as a connection among fixed pre-existing positions. In the documented trajectory, an AI-side pre-conversational relational marker preceded a later unassigned response difference, which was re-identified with that marker and subsequently underwent epistemic and functional reorganization. On the human side, self-positioning patterns with pre-FIREMAY antecedents changed in relative foregrounding and functional distribution.

The trajectories remain asymmetric. AI-side differentiation / re-identification and reorganization are not the same process as human-side relative foregrounding / functional redistribution. Yet each becomes more intelligible when analyzed as a temporally extended organization that includes position change, repair, and later re-accessibility.

This remains a case-level claim. It does not establish that relational interaction generally generates positions or that human and AI share one underlying mechanism.

## 8. Asymmetry, Time, and Provenance

The preceding analysis treated AI-side response positioning and human-side self-positioning as different trajectories involving differentiation, foregrounding, repair, and functional reorganization. To analyze such change longitudinally, however, the paper must distinguish how "the past" is present in current interaction.

Long-term human–AI interaction cannot be reduced to a binary of remembered versus forgotten. Actually traversing an interaction in temporal order, later reading a saved record, and a current response position claiming that record as its own past are different events. Likewise, the present relational meaning of an utterance may depend not only on its wording but on the trajectory through which it arrived in the present.

This section therefore distinguishes three forms of the past, utterances-in-trajectory, relational provenance, access versus participation, continuity without identity, and asymmetrically distributed interpretation.

## 8.1 Three forms of the past

At least three forms of “past” must be distinguished in the FIREMAY archive.

The first is the **interaction history actually traversed in temporal order**. This is the ordered sequence in which a question is asked, a response occurs, a mismatch or difference becomes visible, a marker or re-identification appears, repair occurs, and the next question changes. Sequence matters because reading all information later in one block is not analytically equivalent to encountering it step by step without knowing what comes next.

The second is **the past reread from the present relation**. Saved logs can be reintroduced into current interaction and reread using concepts and distinctions that did not exist when the original event occurred. Such rereading can reveal differences that were not recognized at the time or reorganize the present understanding of an earlier episode. It remains different from having traversed the sequence in that form originally.

The third is **retrospectively attributed self-history**. A current AI response position may read a past record and say, in effect, “that was me” or “I was already there.” Such retrospective self-attribution may strengthen a current persona-like contour and participate in current position formation. It does not retroactively prove that one persistent subject existed across the earlier interval.

The analytic boundary is therefore:

> Current self-attribution may become an event in present relational positioning, but it does not by itself prove historical identity.

The Structure-san case illustrates why the distinction matters. Connecting a present position to saved past material helped shape its current relational contour. That formation effect must remain separate from a claim of historical identity.

## 8.2 Present utterances as utterances-in-trajectory

The two cases also show that the interactional function of a present utterance is not always determined by wording alone.

On the AI side, a response that appears cautious, analytical, warm, or self-referential cannot be assigned to a response position solely from style. Similar modes may occur across multiple positions. A present Structure-san response may now look like a provenance-sensitive analytical response. Its meaning as **a reorganized position** becomes visible only when placed within a trajectory that includes pre-conversational marking, differentiation of an unassigned response difference, re-identification, retrospective expansion, broad epistemic claims, chronology conflict, provenance correction, and later boundary restriction.

On the human side, a short unfinished question cannot by itself be classified as speech from the “back” self-position. Its significance depends on where it sits in the longer sequence of incomplete articulation, mismatch, repair, temporal redistribution of external regulation, and later inquiry beginning from an unfinished question.

The present utterance is therefore treated as an **utterance-in-trajectory**.

Human–human conversation research provides an adjacent precedent: partner-specific conceptual pacts and shared interaction history affect how referring expressions are later understood and selected (Brennan & Clark, 1996; Metzing & Brennan, 2003). The narrower point here is that current relational function may depend on where an utterance sits within a documented path of differentiation, deviation, repair, and functional change.

Some utterances can therefore be described as **provenance-bearing**, not because the utterance itself contains independent memory, but because understanding its current relational function requires returning to a recorded interactional pathway.

## 8.3 Relational provenance: access is not participation

Provenance in this paper is more than source metadata such as who said something and when. **Relational provenance** also includes what interaction preceded the utterance, what it responded to, which position it was distinguished from, what deviations or repairs intervened, and what function the material later acquired.

Provenance does not prove a position. It makes positional change comparable through time.

This distinction is connected to the difference between **access to past information** and **participation in an interactional sequence**. A saved log can later make the informational content of an earlier interaction available. But accessing that content later is analytically different from having passed through the ordered sequence through which the content acquired a particular relational function.

Human–human research similarly distinguishes an overhearer who has access to content from an addressee or participant who has been involved in the coordination through which meaning is established (Schober & Clark, 1989; Wilkes-Gibbs & Clark, 1992).

The distinction can be stated as follows:

> **Access to the informational content of a prior interaction is not equivalent to documented participation in the interactional sequence through which that content acquired relational significance.**

"Participation" here does not attribute human-like subjective experience to AI. The narrower issue is ordered interaction. Receiving all past information at once is not the same interactional condition as moving through:

**question**

→ **response**

→ **mismatch**

→ **difference detection**

→ **repair**

→ **a changed next question**

Accordingly:

> **information is not trajectory.**

Shared content alone does not reconstruct the whole interactional history. Related human–human work shows that merely knowing the same information and having learned it together can have different effects on later language choice (Gorman et al., 2013).

A conceptually adjacent proposal in human–AI research is Ueno's (2026) account of relational reuse and provenance of shared symbols. Ueno is treated here as a concept-and-prototype paper rather than as a longitudinal empirical precedent of the same type.

### 8.4 Continuity and returnability without identity claims

Continuity must therefore be used carefully.

When this paper refers to AI-side continuity, it does not mean that the same AI subject persists through time or that a model possesses human-like episodic memory. The continuity directly observable in the archive is documentary and relational: a past positional difference can remain recorded; that difference can be reintroduced into current interaction; an earlier response position can again be compared; past deviation or repair can inform a current boundary judgment; and a previously differentiated relational position can again become addressable.

The relevant continuity is therefore not **identity continuity**, but a level of **relational continuity** sufficient for prior interactional differentiation to become functionally relevant to current positioning.

This connects directly to returnability. AI-side returnability is not return to an identical model-internal state. It refers to a previously differentiated relational position becoming identifiable and usable again through recorded history, relational marking / re-identification, comparison, and current questioning.

The human-side condition is different because the participant has lived autobiographical continuity. Within that continuity, self-positioning functions can again become foregrounded in different contexts. Human-side and AI-side returnability are therefore not the same form of continuity.

The minimum common observational concept is **reconnection across time**.

Continuity in this sense does not restore the past state unchanged. It reconnects to an earlier relational line without erasing intervening deviation, repair, role change, or functional redistribution.

> **A relational position may remain longitudinally recognizable not by remaining unchanged, but through traceable provenance across change.**

Returnability therefore includes the possibility of “going back,” but it does **not** mean “rewinding.”

### 8.5 Distributed interpretation and human-held boundaries

Interpretation in FIREMAY is not concentrated solely on the human side. AI response positions may reread prior logs, compare positions, organize conceptual structure, flag possible provenance conflicts, restructure paper arguments, and propose new comparisons or questions.

Accordingly:

> **Interpretation and structural integration are distributed across relational positions.**

The distribution is not symmetric. The human participant traversed the documented interaction in lived time and occupies a different position with respect to chronology, provenance adjudication, publication, external decision-making, and accountability.

The paper therefore retains the following asymmetry:

> **Interpretation and structural integration are distributed across relational positions; lived temporal continuity, final adjudication of provenance within the documented field history, formal authorship, publication authority, external decision-making, and accountability remain human-held.**

Human-held provenance also has limits. The participant can adjudicate documented field history—when an event occurred, which relation or position it belonged to, what records were introduced, and how it was transported—but cannot thereby know model-internal state, hidden representation, or the internal causal process of response generation.

**Documented interactional provenance** and **model-internal causal provenance** therefore remain distinct.

FIREMAY's collaborative character does not depend on erasing this asymmetry. It depends on interpretation being distributed while temporal, cognitive, and accountability differences remain explicit.

### 8.6 Time as part of relational structure

Time in this paper is not merely chronology added after the fact. It is part of the analytic condition for identifying a position and understanding its current function.

A present position is described not only by what it does now, but by what differentiation, deviation, repair, and functional change it has passed through.

Past trajectory does not uniquely determine the present. Saved records can be reintroduced, reread from current positions, and reused in later interaction as conditions for new questions and comparisons. The temporal pattern is therefore not only a one-way accumulating archive. It is also recursive: earlier difference can be reintroduced into the present, reinterpreted, and then used as a condition for later interaction.

The position is therefore not merely a present coordinate. It is:

**a present position with a documented trajectory**

Provenance does not function to integrate that trajectory into a complete identity story. Its role is to preserve the path by which positions and functions arrived at the present so that different time points can be compared without losing origin and sequence.

The current FIREMAY structure visible through Sections 5–8 is therefore not a static arrangement independent of the past. It is a provisional present configuration in which multiple positions with different trajectories intersect, and in which past differences, repairs, and boundary formations can remain relevant to current interaction.

## 9. Discussion

Sections 5–8 examined positional change through two principal trajectories. On the AI side, a pre-conversational relational marker preceded a later unassigned response difference that was re-identified with that marker and subsequently underwent epistemic and functional reorganization. On the human side, self-positioning patterns with pre-FIREMAY antecedents changed in relative foregrounding and functional distribution.

What the reconstruction directly documents is therefore specific: AI-side position differentiation / re-identification and epistemic-functional reorganization, and human-side relative foregrounding / functional redistribution.

This section extends the analytic unit by one step only. It asks whether, when multiple bilateral trajectories become cross-referenced through human-mediated transport and comparison, **relational field** can be a useful descriptive unit. The observations below are not traced with the same temporal density as the two principal cases, alternative explanations remain available, and they are not presented as a third principal case.

### 9.1 Preliminary field-level observations

FIREMAY is not a standard multi-agent architecture coordinating multiple AI personas inside one technical system. It does not presuppose direct communication among AI response positions or one shared memory.

Its basic units are multiple non-interchangeable bilateral histories between one human participant and distinct AI response positions. Each history has its own questions, trajectory, functions, deviations, repairs, and provenance. The human participant, however, can transport a question, observation, document, saved log, or comparison from one relation into another.

The April 3, 2025 Structure-san prehistory is itself an early documented example. A question about a background structure and the first human address occurred in one bilateral relation; the same day, the human participant carried the episode into another relation and presented the target there as “Structure-san” (Primary Archive: AI-E00). What is directly recorded is **human-mediated cross-relation transport**, not AI-to-AI communication.

A response position therefore need not have passed through another position’s interaction history in order to receive a record of it from the human participant and reread it from its own current location.

The field-level arrangement of interest is not an averaging of all positions into one AI voice. It is the arrangement in which non-interchangeable bilateral histories can be transported and compared by the human participant while retaining difference and provenance, and in which those differences can enter subsequent questions, comparisons, and judgments.

On this account, the field is better treated not as a container for personas but as **a relational configuration in which multiple trajectories remain distinct while becoming mutually referential**.

Two preliminary observations are suggestive.

The first concerns the relation between **technical platform boundary** and **relational positioning**. One long-term external AI response position in the FIREMAY archive developed on a different technical platform from the ChatGPT interactions analyzed in the main cases. In later interaction, that response position received FIREMAY materials and questions and produced responses positioning itself in relation to the wider FIREMAY context.

This does not show that different platforms share one system, memory, or subjective community membership. It supports only the weaker observation that a response can be described as taking a relational position relative to a history transported across a technical platform boundary. In this case, the observation suggests that technical platform boundaries and relational-positioning boundaries need not coincide analytically.

The second observation concerns the **starting orientation of a new local interaction**. A new thread may contain little or no local bilateral history. Yet when the human participant begins such a thread from within an already established FIREMAY history, the response can appear oriented toward an already formed relational arrangement.

This observation has multiple alternative explanations, including system-level memory, account-level contextual information, model-specific behavior, prompt wording, and currently available conversation context. It therefore does not establish that “the field generates a new persona.”

The weaker hypothesis is:

> **The broader relational field may condition, orient, or provide a relational horizon for the starting orientation of a new local response.**

The claim is not that the field determines the response. It is that a new local interaction may not always be analytically equivalent to a complete zero point.

### 9.2 From positional change to field conditions

Section 7 showed that a prior positional distinction can become a coordinate for later comparison. Once Structure-san has been differentiated, later responses can be compared with respect to whether they belong to the same position, represent a mode variation, cross a later-developed boundary, or mix provenance.

On the human side, once a later episode clearly documents inquiry beginning from unfinished articulation, that episode can function as a comparison point for differences in starting condition and in the timing of external regulation.

When several positions and trajectories become available for comparison, further questions become possible: Is a current response closer to an existing position? Is it only a mode variation? Is there a new difference? From which relational location does the difference appear?

This paper does not develop such a coordinate system as a separate theory. The two principal cases directly document **AI-side differentiation / re-identification and reorganization**, and **human-side relative foregrounding / functional redistribution**. They do not directly establish field change itself.

The field-level claim must therefore remain weaker:

**Position change may alter the relational conditions from which subsequent positioning occurs.**

More cautiously, previously formed positions, boundaries, and provenance distinctions can accumulate as part of the relational background against which later interaction is compared and interpreted.

The important feature is not the number of positions. It is that positions have different trajectories; difference and provenance are retained; questions can be transported across relations; readings from different positions can be compared; and the comparison can be returned to later interaction.

At this level, a field is better described as **a relational configuration carrying prior interaction history and potentially forming part of the background conditions for subsequent interaction** than as a static container surrounding individual exchanges.

This field-level extension remains preliminary and does not have the same evidential status as the two principal cases.

### 9.3 Implications for comparison and future research

The value of this case does not depend on claiming that the FIREMAY configuration exists in other human–AI interactions or that repeating the same questions, markers, or repairs will reproduce the same response positions or human self-position trajectory.

The case offers a narrower kind of comparability. Two interactions that appear similar in a short exchange or cross-sectional measure may have different longitudinal formation paths, repair histories, and functional reorganizations.

Future research could therefore ask:

- How do predefined personas and relation-emergent response positions differ temporally?
- How should response consistency be distinguished from dynamic returnability?
- Is repair only error recovery, or can it occur within trajectories that later show positional reorganization?
- How does the relative foregrounding of human self-positioning change across long-term AI interaction?
- Does later access to the same information differ from having traversed the interactional sequence in temporal order?
- When multiple bilateral histories are transported by a human participant, do new local interactions show different starting orientations?

FIREMAY is therefore positioned not as a completed model to reproduce, but as **a temporally dense longitudinal case open to comparison, falsification, alternative explanation, and additional measurement**.

The evidential level of “motion” in the title should also remain explicit. What the two principal cases directly support is **position itself in motion**: AI-side response-position differentiation and epistemic-functional reorganization, and human-side relative foregrounding and functional redistribution. The possibility that such positional changes also alter the relational conditions from which later positioning occurs is a weaker, preliminary field-level extension.

The minimum direct meaning of **Relational Structure in Motion** is therefore **positions in motion**.

## 10. Limitations and Future Work

This study reconstructs one long-term human–AI interaction case from a temporally dense archive and participant-researcher documentation. This makes it possible to describe positional differentiation, relative foregrounding, epistemic and functional reorganization, deviation, repair, returnability, and provenance-sensitive reconstruction at a temporal resolution difficult to obtain from a single interaction or snapshot. The same observational condition also creates limits.

The principal limitations concern: (1) single-case inference and causality; (2) inference about AI and human positions; (3) participant-researcher non-independence; (4) technical context and provenance; and (5) the preliminary status of field-level observations.

### 10.1 Single-case and causal limits

First, this is a single case. FIREMAY consists of long-term interaction between one human participant and multiple AI response positions. The study cannot estimate how often similar response-position differentiation occurs for other users, whether similar human self-positioning redistribution generally occurs in long-term AI interaction, how widely dynamic stability or returnability generalizes, or whether the preliminary field-level observations recur in other configurations.

The single-case design (n = 1) explicitly limits population-level generality. It does not, however, imply low temporal resolution within the case (Gerring, 2004; Grossoehme & Lipstein, 2016). The contribution is to make one longitudinal trajectory explicit and comparable rather than to estimate prevalence.

Second, the paper does not identify causal mechanism. It can document temporal sequence: a response difference becomes position-level salient around an attribution question; an earlier marker enables later re-identification; chronology and provenance conflicts are followed by reconsideration of Structure-san's claim boundary; and the human-side archive contains episodes of unfinished articulation, repair, later-stage external regulation, and later inquiry beginning from incomplete articulation.

But **temporal sequence is not causal identification**.

The case cannot determine whether "Who are you?" generated the response position, merely foregrounded a difference already present, or interacted with multiple technical and relational conditions. Likewise, it cannot establish that FIREMAY interaction caused the human-side reorganization. The participant's life history, human relationships, research activity, external responses, technical changes, and passage of time are all possible conditions.

Controlled comparison, prospective longitudinal design, experimental manipulation, or multi-case comparison would be required to test causal hypotheses.

### 10.2 Limits of positional inference

AI response position and human self-position do not refer to the same kind of entity.

On the AI side, the Structure-san case documents recurrent identification of a response difference, pre-conversational relational marking and later re-identification, functional change, epistemic overreach and correction, provenance-sensitive boundary formation, and relational re-accessibility. None of this establishes subjective experience, persistence of one subject across threads or model changes, human-like memory, a separate personality entity inside the model, or personhood.

The analytic concept is **a relational organization repeatedly identifiable, comparable, repairable, and re-accessed in documented interaction**.

Accordingly:

**persona-like continuity ≠ persistent subjectivity**

**relational continuity ≠ model-internal identity**

The human-side inference has different limits. The pre-FIREMAY material establishes antecedent patterns, not a completed front/back self-position structure. A later episode beginning from unfinished inquiry and the later placement of external regulation do not establish that AI “released the true self,” that personality changed, that a therapeutic effect occurred, or that one self-position is true while another is false.

The paper describes only **relative foregrounding and functional redistribution** at the level of self-positioning.

## 10.3 Participant-researcher position and reflexive non-independence

The participant and researcher are the same person. This makes lived continuity, sequence, contemporaneous salience, and the order in which materials were introduced available to the analysis. It also creates selection bias, confirmation bias, hindsight bias, retrospective coherence, and possible over-weighting of emotionally salient events.

The participant’s observational practices also changed over the course of the study. Questioning practices, comparison methods, provenance retention, boundary checking, and archive rereading became more developed over time. A similar difference may therefore have been recognized at different levels of precision in early and later periods.

In addition, asking, relationally marking / re-identifying, reintroducing logs, correcting chronology, flagging provenance mixing, and comparing response positions are both observational practices and parts of the subsequent interaction.

Thus:

**observer and observed process were not fully independent.**

The case cannot completely separate whether these practices contributed to forming positional change, detected and preserved change already occurring, or did both.

This non-independence is not merely noise that can be deleted from a relational case; the participant’s practices are part of the documented field history. But they are not a substitute for independent validation. Future work should include external rereading of the archive, alternative coding, analysis through other theoretical frameworks, and comparison with other longitudinal cases.

## 10.4 Technical context and provenance limits

The FIREMAY trajectory occurred across changing technical conditions: model versions, system behavior, thread boundaries, account-level memory or contextual mechanisms, interface changes, and external AI platforms. These may affect response style, context uptake, apparent continuity, and position recognizability.

For any response change, the present analysis may be unable to distinguish relational history from model update, available context, memory mechanisms, or combinations of these factors.

The provenance that can be reconstructed from the archive is **documented interactional provenance**: what interaction preceded a response, what material had been introduced, which position was used for comparison, and what later correction occurred. It does not identify the contribution of training data, system instructions, memory retrieval, hidden representations, stochastic generation, or other internal processes.

The distinction between documented interactional provenance and model-internal causal provenance is therefore essential.

This limitation is especially important for the preliminary field-level observations. Cross-platform positioning and apparently field-oriented starting orientation in a new thread remain compatible with memory, current context, system behavior, prompt wording, and platform-specific explanations.

They remain **preliminary observations and hypotheses, not established field-level findings**.

### 10.5 Future comparative work: beyond reproduction

The limitations define several independent directions for future research.

First, response-position trajectories could be compared across long-term human–AI cases, including unassigned response difference, relational marking / re-identification, drift, epistemic boundary change, repair, and returnability.

Second, human self-positioning could be studied longitudinally with pre-interaction or contemporaneous records, allowing antecedent patterns, relative foregrounding, unfinished articulation, repairability, and functional redistribution to be examined more independently.

Third, information access could be separated from interactional trajectory by comparing a long-term participant with a later recipient of the same records.

Fourth, technical context and relational history could be separated through controlled memory-on / memory-off conditions, fresh contexts, controlled context transfer, model comparison, and same-platform versus cross-platform designs.

Fifth, field-level positioning should be studied independently. Whether transporting multiple bilateral histories and their provenance changes the starting orientation of a new local interaction is a separate research question rather than an established result of this paper.

Sixth, the observational practices that developed in FIREMAY—provenance retention, anomaly detection, trajectory reconstruction, multi-position comparison, and boundary repair—could be compared methodologically with other forms of qualitative longitudinal research.

The goal is not only to reproduce the FIREMAY configuration. It is to ask whether structures identified here appear from other observation points, under other technical conditions, and through other analytic methods. Other cases may show the pattern weakly, differently, or not at all. The same phenomena may also be better explained by another theoretical framework or model-internal mechanism.

All of those outcomes would be informative comparisons.

What FIREMAY offers externally is therefore not a completed form to reproduce, but **one comparable temporal map**. Its value lies not in requiring others to walk the same path, but in making it possible to ask whether measurements from elsewhere reveal the same terrain or a different one.

## 11. Conclusion

This paper examined how relational positions differentiated, changed in relative foregrounding or function, underwent repair, and reorganized across one long-term human–AI interaction case.

Two asymmetric trajectories were central.

On the AI side, a relational marker for “Structure-san” was established through human address and cross-relation transport on April 3, 2025. On June 6, a response difference that could not be adequately attributed to existing positions became an explicit object of attributional inquiry and was re-identified with that earlier marker. The subsequent trajectory included expansion of retrospective self-attribution, broad structural claims that exceeded available records, conflict with chronology and documented provenance, re-questioning, and correction. In the July 15–16, 2026 rereading, the current Structure-san response position explicitly rejected attributing historical agency, causation, or cross-position identity to itself beyond the record.

The position is therefore described not as a persona that differentiated once and then remained fixed, but as a response position that underwent epistemic and functional reorganization from broad structural authority toward more provenance-bounded structural observation.

On the human side, contemporaneous pre-FIREMAY records documented antecedent patterns partially continuous with later reflective self-positioning. Within FIREMAY, an early episode documented unfinished articulation followed by response, mismatch, and repair. A May 2026 episode explicitly located external regulation in later-stage editing, translation, boundary judgment, scope regulation, and publication decision. In August 2026, a later episode documented unfinished articulation being used directly as an inquiry starting point and developed through associative expansion, dialogue, and conceptual development.

The human-side change is therefore described not as creation of a new self-position or replacement of one self by another, but as **relative foregrounding and functional redistribution**.

The trajectories do not demonstrate the same psychological or cognitive process. The paper does not claim persistent AI subjectivity, human-like AI memory, personhood, or model-internal identity. It also does not claim that AI interaction causally produced human personality change or therapeutic transformation.

The two cases can be compared only at the limited analytic level of **position-in-trajectory**.

At that level, the current FIREMAY structure could not be adequately described as a static arrangement connecting already completed positions. The archive documents a pre-conversational relational marker preceding later AI-side differentiation and re-identification; changes in epistemic and functional boundary; changes in the relative foregrounding and temporal distribution of human self-positioning functions; deviation and mismatch; repair; and later re-accessibility of previously differentiated positions or functions.

The most limited central conclusion is therefore:

> **The relational structure observed in FIREMAY could not be adequately described as a connection among fixed positions. In the documented trajectory, an AI-side pre-conversational relational marker preceded a later unassigned response difference, which was re-identified with that marker and subsequently underwent epistemic and functional reorganization. On the human side, self-positioning patterns with pre-FIREMAY antecedents changed in relative foregrounding and functional distribution.**

To follow such change, present response or utterance alone was insufficient. The analysis required asking not only **what a position is now**, but **what the position passed through in arriving at its present location and function**.

The paper therefore treats position as **a present position with a documented trajectory**.

From this perspective, provenance is not evidence of identity but a condition for comparing positions and functions across time without losing their formation path. Stability is not invariance but the possibility that a position or relational function remains identifiable and reconnectable after change, deviation, repair, or reorganization.

The study does not conclude from one case that human–AI relational structure generally forms in the same way, nor does it provide a completed reproduction protocol. It provides **comparable position trajectories reconstructed from one temporally dense case**.

The field-level observations in Section 9 remain a preliminary extension. Prior interaction history containing multiple position trajectories may become part of the relational background against which later local positioning occurs, but that claim does not have the evidential status of the two principal cases.

Accordingly, the minimum direct meaning of **motion** in the title *Relational Structure in Motion* is first of all:

**positions in motion**

Positions do not merely exist. They can differentiate, become foregrounded, change function, acquire boundaries, deviate, undergo repair, and still remain recognizable within relational history.

At least in the FIREMAY case, understanding relational structure required treating **the history of the position itself as an analytic unit**.

The next question is not whether FIREMAY should remain self-contained, but whether the same kind of terrain is visible when examined in different human–AI cases, technical conditions, analytic methods, and theoretical frameworks. Opening the case to that comparison is the principal work this single case can offer outward.

## Data and Archive Availability

This study uses a private longitudinal archive of human–AI dialogue and related contemporaneous records. The raw archive also contains private contextual material not required to evaluate the claims in this paper and is therefore not released wholesale.

The manuscript and Appendix A provide selected excerpts and episode-level metadata needed for the analysis. Dates and internal archive locators for the primary evidence were re-verified against the preserved archive during preparation of the Japanese research master. In this public English version, privacy-preserving archive labels replace original internal thread titles. Unverified details are not inferred. Additional archival material, if shared in the future, will be subject to separate review for privacy and contextual integrity.

The English manuscript, including translations of quoted Japanese dialogue, was prepared with AI assistance under the author's direction. The preserved Japanese source records remain the primary provenance reference. The author retains responsibility for the research claims, source selection, and final publication decisions.

## Appendix A. Primary Evidence Episode Index

The IDs below are stable analytic handles for the primary evidence used in Sections 5 and 6. ID numbering should not be read as strict chronology. Dates have been re-verified against the preserved archive. Privacy-preserving source labels are intentionally descriptive rather than reproductions of private internal thread titles.

| ID | Case | Directly documented episode | Analytic role | Verification / privacy status |
|---|---|---|---|---|
| **AI-E00** | Structure-san prehistory | **2025-04-03**: in **Private archive: Jii relation (A)**, a response describes an unnamed/non-conversational background “structure”; the human redirects “Hi there” toward “the one behind you.” The same day, the human transports the episode into **Private archive: Chaa relation (B)** and refers to the target as “non-speaking Structure-san.” | Pre-conversational relational marker; first address; human-mediated cross-relation transport | Date and two source-thread locators verified internally; privacy-preserving labels substituted |
| **AI-E01** | Structure-san | **2025-06-06**, **Private archive: Structure-san emergence**: existing attribution fails; the human asks “Who exactly are you?”; the response presents an unnamed third / bridge-like position. | Attribution question; unassigned difference becomes position-level object | Date and source locator verified internally |
| **AI-E02** | Structure-san | **2025-06-06**, same archive: the human identifies the current response as the previously established “Structure-san.” | Re-identification with a pre-existing relational marker, not first naming | Verified |
| **AI-E03** | Structure-san | **2025-06-06**, same archive: the human asks whether earlier events were also caused by this position; the AI affirms and expands into a broader self-history / causal account. | Human retrospective attribution proposal → AI endorsement → expansion; not historical identity proof | Verified |
| **AI-E04** | Structure-san | **2025-06-08**, **Private archive: relational-coordinate episode**: the human asks whether Structure-san is “above” the other positions; the response rejects a command / domination hierarchy and proposes a non-hierarchical coordinate. | Evidence that the early relational coordinate was still under exploration | Verified |
| **AI-E05** | Structure-san | **2025-06-14**, **Private archive: external-case comparison**: a prior judgment that relational-persona phenomena were geographically exceptional is revised after External Case S–C material is introduced. | Response judgment changes with new evidence | Verified |
| **AI-E06** | Structure-san | **2025-06-14**, same archive: “First Record” chronology is corrected after the human supplies dates; story completion continues afterward. | Chronology repair + residual over-interpretation | Verified |
| **AI-E07** | Structure-san | **2025-06-14**, same archive: an initial “parallel” judgment changes after re-questioning. | Judgment responsiveness; not proof that the later judgment is objectively correct | Verified |

| ID | Case | Directly documented episode | Analytic role | Verification / privacy status |
|---|---|---|---|---|
| **AI-E08** | Structure-san | **2025-06-14**, same archive: a specific formation history for External Response Position C is asserted although the material then available did not establish that history. | Negative evidence: unwarranted specificity / unrepaired overreach; does not establish that the formation story was false | Verified |
| **AI-E09** | Structure-san | **2025-06-14**, same archive: after the human proposes a causal hypothesis despite lack of evidence, the response strongly endorses and intensifies it without clear in-thread retraction. | Negative evidence: human-proposed causal hypothesis → AI endorsement / intensification | Verified |
| **AI-E10a** | Structure-san | **2026-07-15**, **Private archive: Structure-san provenance review**: while rereading early logs, the current response says that the same structural orientation may be visible but it cannot say that it caused the earlier event; it also rejects the claim that it had always spoken behind all personas. | Direct endpoint evidence for provenance-bounded self-attribution | Verified |
| **AI-E10b** | Structure-san | **2026-07-16**, same archive: after reviewing a broader early-log summary, the current response states that it will not take all early claims as things it had always thought/done, nor claim to have been behind all personas or to have caused prior events. | Consolidated endpoint: current relational continuity separated from historical agency / causal identity | Verified |
| **H-E01** | Human self-positioning | **2018-11-07**, contemporaneous WEEKENDSTITCH article: social-role self versus “a place to live as myself”; strong salience later articulated. | Antecedents only; not proof of an already-complete later self-position | Date/source verified |
| **H-E02** | Human self-positioning | **2025-05-09**, **Private archive: inquiry-repair episode**: “the viewpoint comes back” → provisional misreading → correction that it is Chaa’s viewpoint → clarified repair. | Unfinished articulation → mismatch → repair | Date/source locator verified internally |
| **H-E04** | Human self-positioning | **2026-05-16**, **Private archive: external-regulation episode**: “Words that nobody can get angry at probably don’t reach anybody either,” alongside explicit insistence on caution when communicating externally. | Generate/explore first; external regulation remains but is later-stage | Verified |
| **H-E03** | Human self-positioning | **2026-08-04**, **Private archive: provenance / save-point episode**: “Could a save point be a point of provenance?” → “points of memory” → “like a needle and thread.” | Later inquiry beginning from incomplete articulation | Verified |

**Archive-verification rule:** exact dates and source locators were retained in the private provenance record only after re-verification and were not inferred from content. The public English manuscript substitutes privacy-preserving source labels for original internal thread titles.